\documentclass{article}

\usepackage[english]{babel}

\usepackage[letterpaper,top=2cm,bottom=2cm,left=3cm,right=3cm,marginparwidth=1.75cm]{geometry}

\usepackage{multirow}
\usepackage{amsmath,amssymb,amsthm}
\theoremstyle{definition} 
\usepackage{relsize}
\usepackage{graphicx}
\usepackage{listings}
\usepackage{xcolor}
\usepackage{booktabs}
\usepackage{fancyref}
\usepackage[parfill]{parskip}
\usepackage{todonotes}
\usepackage{import}
\usepackage{dsfont}
\usepackage{wrapfig} 
\usepackage{lineno}
\usepackage{float}
\usepackage{caption}
\usepackage[skip=10pt]{subcaption}
\usepackage{bm}
\usepackage{bbm}
\usepackage{commath}
\usepackage[section]{placeins}
\usepackage{fancyhdr}
\usepackage{natbib}
\usepackage[nodayofweek]{datetime}

\usepackage{algorithm}
\usepackage{algorithmic}
\mathchardef\mhyphen="2D 

\DeclareMathOperator*{\argmin}{argmin}
\DeclareMathOperator*{\argmax}{argmax}

\def\RR{\mathbb{R}}

\newcommand{\EE}[1]{\mathbb{E}\left[#1\right]}

\newcommand{\VV}[1]{\mathbb{V}\left[#1\right]}

\numberwithin{equation}{section}

\usepackage{amsmath}
\usepackage{tikz}
\usetikzlibrary{calc}
\usetikzlibrary{matrix}
\usetikzlibrary{chains}
\usetikzlibrary{positioning}
\usetikzlibrary{decorations.pathreplacing}
\usetikzlibrary{arrows}
\usetikzlibrary{bayesnet}

\newtheorem{theorem}{Theorem}[section]
\usepackage{graphicx}
\usepackage[colorlinks=true, allcolors=blue]{hyperref}

\newcommand{\myref}[2]{\hyperref[#2]{#1 \ref*{#2}}}
\newcommand\bigCI{\mathop{\underline{\raisebox{0pt}[0pt][1pt]{$\;||\;$}}}}

\title{Filling in Missing FX Implied Volatilities with Uncertainties: \newline Improving VAE-Based Volatility Imputation}
\author{
    Achintya Gopal \\
    Bloomberg \\
    New York, NY, USA \\ 
    \texttt{agopal6@bloomberg.net}
}

\begin{document}
\maketitle

\begin{abstract}
Missing data is a common problem in finance and often requires methods to fill in the gaps, or in other words, imputation. In this work, we focused on the imputation of missing implied volatilities for FX options. Prior work has used variational autoencoders (VAEs), a neural network-based approach, to solve this problem; however, using stronger classical baselines such as Heston with jumps can significantly outperform their results. We show that simple modifications to the architecture of the VAE lead to significant imputation performance improvements (e.g., in low missingness regimes, nearly cutting the error by half), removing the necessity of using $\beta$-VAEs. Further, we modify the VAE imputation algorithm in order to better handle the uncertainty in data, as well as to obtain accurate uncertainty estimates around imputed values.
\end{abstract}

\section{Introduction}

Missing data is a prevalent problem throughout finance and often requires methods to fill in the gaps, or, in other words, imputation. Examples of missing data in finance include reporting inconsistencies (e.g.,  environmental, social, and governance (ESG) data (e.g., \cite{ESGImp2020})) and a lack of pricing data due to illiquidity such as imputing options volatility surfaces  (e.g., \cite{Heston1993ACS,Dupire1994LocalVol,Brigo2002MLN,bergeron2022variational,richert2022vaeswaption}) and imputing yield curves  (e.g., \cite{vasicek1982spline,nelsonsiegel1987,svensson1994nss,sokol2022amm}). In this work, we will specifically focus on the task of imputing missing implied volatilities for FX options.

One classical method to fill in missing values in options surfaces is to fit dynamics models (e.g., Black Scholes, Heston \citep{Heston1993ACS}, local volatility \citep{Dupire1994LocalVol}) that are consistent with the observed implied volatilities. Another set of methods relies on fitting static models that provably are no-arbitrage (e.g., Mixture of Log Normals \citep{Brigo2002MLN}, SSVI \citep{gatheral2014ssvi}). 

While these methods, that are often used in practice, perform well, and are interpretable, given the popularity of neural networks, a natural question is to ask whether or not neural networks can perform better, even if at the cost of interpretability? The value of neural networks is that they are both universal approximators \citep{cybenko1989approximation} and are able to generalize well to unseen data. 

One neural network-based method to imputing missing data that has been used recently in the financial literature is variational autoencoders (VAEs) \citep{bergeron2022variational,richert2022vaeswaption}. While only recently explored in the context of finance, using VAEs for imputing data has been explored for many years within the machine learning literature \citep{rezende2014stochastic,mattei2018dlvm,mattei2019miwae}. However, while the point estimates that come from the VAEs have been tested in terms of imputing FX volatility surfaces, the uncertainty component has not been fully explored. 

In this work, we change the commonly used imputation methodology used in VAEs to include uncertainty estimation (distributional imputation), since the value of VAEs in uncertainty modeling. In other words, we focus not only on point estimates, but also on $p(\mathbf{x}_{missing} | \mathbf{x}_{obs})$.
Additionally, to improve the imputation of FX volatility surfaces, we:
\begin{itemize}
    \item incorporate heteroscedastic noise into the decoder/generator \citep{nazabal2018mvae}, and
    \item replace the normal multilayer perceptron (MLP) architecture with residual networks \citep{he2016resnet,Gorishniy2021TabRes}.
\end{itemize}
Finally, we show that, when tested against many classical methods for imputing FX volatility surfaces, VAEs are unable to outperform them without changing the architecture to residual networks.

\section{Related Work}\label{sec:related_work}

One method to impute missing data that has been used recently in the financial literature is variational autoencoders (VAEs). \citet{bergeron2022variational} and \citet{sokol2022amm} train VAEs directly on market data and then generate point estimates by finding the value in the latent space that best reconstructs the observed data. \citet{richert2022vaeswaption} enhances the imputation methodology by using pseudo-Gibbs sampling, originally introduced by \citet{rezende2014stochastic}. More details of these approaches are given in \myref{Section}{sec:comparison_vae_imp}.

While only recently explored in the context of finance, using VAEs for imputing data has been explored for many years within the machine learning literature. \citet{rezende2014stochastic} introduced a Gibbs-influenced imputation method; however, this method required that the encoder (introduced later in \myref{Section}{sec:vi}) is an accurate estimate of the posterior. \citet{mattei2018dlvm} corrected for this by introducing a Metropolis-within-Gibbs extension. However, this will converge to the right distribution at an impractical convergence rate \citep{mattei2019miwae}. \citet{mattei2019miwae} replaced the Metropolis-within-Gibbs algorithm with importance weighting, an approach that is explained in more detail in \myref{Section}{sec:imputation}.

\section{Background: VAE}

The crux of this work is the usage of variational autoencoders (VAEs); in this section, we review the model and discuss some common misconceptions of the model. 

Often VAEs are described as ``autoencoders with a KL term to regularize the latent encoding''. While this gives a simple description of the loss function, we believe it obfuscates the derivation, usage, and extensions of VAEs. A good description informs users what components can be modified. With the usual description, the method of regularization seems to be a natural way to modify the model; in other words, the usual description positions VAEs as a way to learn a structured latent embedding space. However, in our opinion, a VAE is most useful for imputation when viewed from the generative and statistical angle. To this end, we prefer to describe VAEs as ``\textit{a deep latent variable model (DLVM) that has been trained using variational inference}.'' We will define both DLVM (\myref{Section}{sec:dlvm}) and variational inference (\myref{Section}{sec:vi}) in more detail later in this section. We argue that viewing neural networks in this manner allows for a more holistic approach to both fitting the model and imputing data (\myref{Section}{sec:imputation}).

\subsection{Deep Latent Variable Model (DLVM)}\label{sec:dlvm}

What is a deep latent variable model (DLVM)? Statistical models are often easiest to explain via the generative process; thus, a latent variable model can generically be described as:
\begin{align*}
    \mathbf{z} &\sim p(\mathbf{z})
\\ \mathbf{x} &\sim p(\mathbf{x} | \mathbf{z})
\end{align*}
where $\mathbf{x}$ is observed and $\mathbf{z}$ is unobserved (latent).
A \textit{deep} latent variable model uses a neural network to parameterize the distributions $p(\mathbf{z})$ and $p(\mathbf{x} | \mathbf{z})$. 

The most common way this is achieved is:
\begin{align}
\begin{split}
    \mathbf{z} &\sim \mathcal{N}(0, I_d)
\\ \mathbf{x} &\sim \mathcal{N}(\mu_\theta(\mathbf{z)}, \sigma_x) \label{eqn:vanilla_gen}
\end{split}
\end{align}
where $\mathbf{z}\in\RR^{d}$ ($d$ will be referred to as latent size or embedding size), $I_d$ is an $d \times d$ identity matrix,  $\mathbf{x}\in\RR^{p}$ ($p$ is the dimensionality of the data, or, in other words, the number of features present in the data), and $\sigma_x \in \RR$ (we will relax the assumption of $\sigma_x$ being a scalar later in \myref{Section}{sec:hetero}).
We assume $p(\mathbf{z})$ is a standard Normal distribution and 
$p(\mathbf{x} | \mathbf{z})$ is a conditional Normal distribution where the mean is parameterized by a neural network $\mu_\theta(\mathbf{z)}$ (where $\theta$ denotes the parameters of the function $\mu_\theta$) and the standard deviation is a scalar independent of $\mathbf{z}$ ($\sigma_x$).
The standard deviation either can be fixed throughout training, or can be learned  through an unconstrained parameter $s$ where $\sigma_x = e^s$ or $\sigma_x = \log(1 + e^s) = \text{Softplus}(s)$. We will revisit this comment in \myref{Section}{sec:beta_vae} in the context of Beta-VAEs.

In this parameterization, we point out one common mistake when sampling from VAEs: sampling from a VAE requires \textit{two sampling steps}. Often, when sampling from a VAE, practitioners will sample from $p(\mathbf{z})$ (a Normal distribution) and then use $\mu_\theta(\mathbf{z})$ as a sample from the model. However, this is not the true generative process by construction in \myref{Equation}{eqn:vanilla_gen}. To sample $\mathbf{x}$, we must further sample from another Normal centered at $\mu_\theta(\mathbf{z})$ with noise $\sigma$. In \myref{Appendix}{sec:sampling_noise}, using a simple example, we show the negative results of not accounting for this second sampling step.

When fitting distributions to data, the most common method is maximum likelihood estimation (MLE):
$$ \argmax_{\theta} \sum_{i=1}^{N} \log p(\mathbf{x_i} | \theta) $$
where $\{\mathbf{x_i}\}_{i=1}^{N}$ denotes the data we want to fit our model on. For a latent variable model, the log-likelihood is:
 $$ \log p(\mathbf{x_i} | \theta) = \log \int p(\mathbf{x_i} | \mathbf{z}, \theta) p(\mathbf{z}) d\mathbf{z}$$
This integral is the reason we cannot directly perform gradient descent of the total log-likelihood; we instead use \textit{variational inference}.

\subsection{Variational Inference (VI)}\label{sec:vi}

Variational inference approximates the log-likelihood $\log p(\mathbf{x_i} | \theta)$ by approximating $p(\mathbf{z} | \mathbf{x_i})$ (also known as the posterior). If we were to know $p(\mathbf{z} | \mathbf{x_i})$, we could approximate the log integral via:
$$ \log p(\mathbf{x_i} | \theta) = \mathbb{E}_{\mathbf{z} \sim p(\mathbf{z} | \mathbf{x_i})}\left[\log  \frac{p(\mathbf{x_i} | \mathbf{z}, \theta) p(\mathbf{z})}{p(\mathbf{z} | \mathbf{x_i})}\right] $$
However, even though we cannot compute $p(\mathbf{z} | \mathbf{x_i})$ exactly, we can approximate it with a variational distribution $q(\mathbf{z} | \mathbf{x_i})$. Using this approximation, we get the following lower bound:
\begin{align}
    \log p(\mathbf{x_i} | \theta) &= \log \int p(\mathbf{x_i} | \mathbf{z}, \theta) p(\mathbf{z}) d\mathbf{z}
\\ &= \log \int p(\mathbf{x_i} | \mathbf{z}, \theta) p(\mathbf{z}) \frac{q(\mathbf{z} | \mathbf{x_i})}{q(\mathbf{z} | \mathbf{x_i})} d\mathbf{z}
\\ &\geq \mathbb{E}_{\mathbf{z} \sim q(\mathbf{z} | \mathbf{x_i})}\left[\log  \frac{p(\mathbf{x_i} | \mathbf{z}, \theta) p(\mathbf{z})}{q(\mathbf{z} | \mathbf{x_i})}\right]
\\ &= \mathbb{E}_{\mathbf{z} \sim q(\mathbf{z} | \mathbf{x_i})}\left[\log  p(\mathbf{x_i} | \mathbf{z}, \theta)\right] + \text{KL}\left(q(\mathbf{z} | \mathbf{x_i})\ ||\ p(\mathbf{z})\right)
\\ &=\text{ELBO}(\mathbf{x_i} | \theta, q) = -\mathcal{L}_{\text{VAE}}(\mathbf{x_i}; \theta, \phi)
\end{align}
where the last step comes from using Jensen's inequality. We depict the VAE loss in \myref{Figure}{fig:vae_diagram}.
A useful property of the ELBO (evidence lower bound) is that the gap between it and the true log-likelihood is the KL divergence between the variational distribution and true posterior:
$$ \log p(\mathbf{x_i} | \theta) - \text{ELBO}(\mathbf{x_i} | \theta, q) = \text{KL}\left(q(\mathbf{z} | \mathbf{x_i}, \theta) \ ||\ p(\mathbf{z} | \mathbf{x_i}, \theta)\right) $$

Analyzing the $\text{ELBO}(\mathbf{x_i} | \theta, q)$, we can see that the description of it as an autoencoder comes from the first term, i.e., $\mathbb{E}_{\mathbf{z} \sim q(\mathbf{z} | \mathbf{x_i})}\left[\log  p(\mathbf{x_i} | \mathbf{z}, \theta)\right]$. To make it clearer, plugging in the log-likelihood of a Normal distribution as defined in generative process (\myref{Equation}{eqn:vanilla_gen}), we get:
\begin{equation}\label{eqn:normal_ll}
 \mathcal{L}_{\text{VAE}}(\mathbf{x_i}; \theta, \phi) = \mathbb{E}_{\mathbf{z} \sim q(\mathbf{z} | \mathbf{x_i})}\left[\frac{1}{2}\norm{\frac{\mathbf{\mathbf{x_i}} - \mu_\theta(\mathbf{z})}{2 \sigma_x }}^2_2 + \log \left(\sigma_x d \sqrt{2 \pi}\right) \right] - \text{KL}\left(q(\mathbf{\mathbf{z}} | \mathbf{\mathbf{x_i}}) || p(\mathbf{\mathbf{z}})\right)
\end{equation}
The first term is similar to an autoencoder with a mean squared error reconstruction term.

\begin{figure}
\begin{center}
\begin{tikzpicture}[
  every neuron/.style={
    circle,
    minimum size=0.3cm,
    thick
  },
  every data/.style={
    rectangle,
    minimum size=0.4cm,
    thick
  },
]

  \node [align=center,data 1/.try, minimum width=1.5cm, draw] (input-x)  at ($ (2.5, 0) $) {$\mathbf{x}$};

  \node [align=center,data 1/.try, minimum width=1.5cm, draw] (q-mu)  at ($(input-x) + (-0.9,1.5)$) {$\mathbf{\mu_{z|x}}$};
  \node [align=center,data 1/.try, minimum width=1.5cm, draw] (q-sigma)  at ($(input-x) + (0.9,1.5)$) {$\mathbf{\sigma_{z|x}}$};

  \node [align=center,data 1/.try, minimum width=1.5cm] (enc-desc)  at ($(input-x) + (-3.0,0.75)$) {Encoder (Inference) \\ $q_\phi(z|x)$};

  \node [align=center,data 1/.try, minimum width=1.5cm] (enc-desc)  at ($(input-x) + (3.5, 2.0)$) {$KL\left(q_\phi(z | x_i)\ ||\ p(z)\right)$};

  \node [align=center,data 1/.try, minimum width=1.0cm] (z)  at ($(input-x) + (0.0,2.5)$) {Sample from $\mathcal{N}(\mu_{z|x}, \sigma_{z|x})$};
  \node [align=center,data 1/.try, minimum width=1.5cm, draw] (input-z)  at ($(input-x) + (0.0,3.0)$) {$\mathbf{z}$};

  \node [align=center,data 1/.try, minimum width=1.5cm, draw] (p-mu)  at ($(input-z) + (-0.9,1.5)$) {$\mathbf{\mu_{x|z}}$};
  \node [align=center,data 1/.try, minimum width=1.5cm, draw] (p-sigma)  at ($(input-z) + (0.9,1.5)$) {$\mathbf{\sigma_{x|z}}$};
  \node [align=center,data 1/.try, minimum width=1.5cm] (dec-desc)  at ($(input-z) + (-3.0,0.75)$) {Decoder  (Generator) \\ $p_\theta(x|z)$};

\node [align=center,data 1/.try, minimum width=1.0cm] (x)  at ($(input-z) + (0.0,2.5)$) {Sample from $\mathcal{N}(\mu_{x|z}, \sigma_{x|z})$};
  \node [align=center,data 1/.try, minimum width=1.0cm] (px)  at ($(input-z) + (3.2,2.0)$) {$\log  p_\theta(x_i | z)$};
  \node [align=center,data 1/.try, minimum width=1.5cm, draw] (output-x)  at ($(input-z) + (0.0,3.0)$) {$\mathbf{\hat{x}}$};

  \draw [black,solid,->] ($(input-x.north)$) -- ($(q-mu.south)$);
  \draw [black,solid,->] ($(input-x.north)$) -- ($(q-sigma.south)$);

  \draw [black,solid,->] ($(q-mu.north)$) -- ($(z.south)$);
  \draw [black,solid,->] ($(q-sigma.north)$) -- ($(z.south)$);

  \draw [black,solid,->] ($(input-z.north)$) -- ($(p-mu.south)$);
  \draw [black,solid,->] ($(input-z.north)$) -- ($(p-sigma.south)$);

  \draw [black,solid,->] ($(p-mu.north)$) -- ($(x.south)$);
  \draw [black,solid,->] ($(p-sigma.north)$) -- ($(x.south)$);

\end{tikzpicture}
\end{center}
\caption{A diagrammatic representation of a vanilla VAE. The two terms on the right side of the diagram are the two terms of the loss function. Interestingly, the KL term is independent of $\theta$ and hence can be calculated before running the decoder (generator) network. }\label{fig:vae_diagram}
\end{figure}
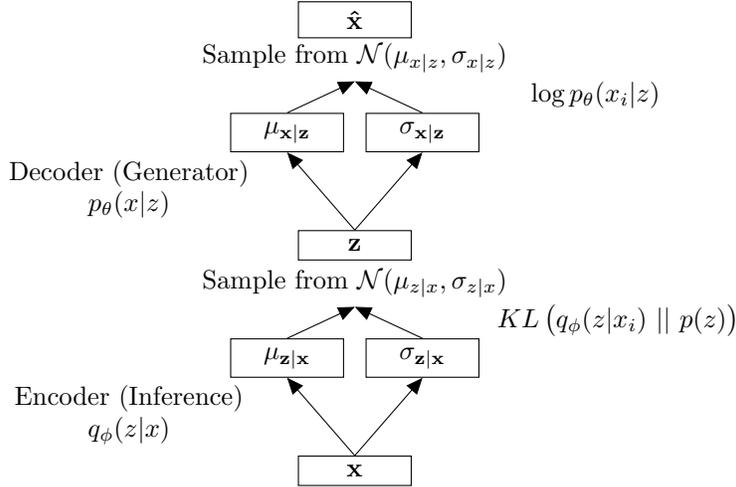

\subsection{Reparameterization Trick}
To train a VAE, we parametrize $q(\mathbf{z} | \mathbf{x_i})$ with a neural network, often as:
\begin{align}
    \mathbf{h} &= f_\phi(\mathbf{x_i})
\\ \mathbf{z}  &\sim \mathcal{N}(\mu_\phi(\mathbf{h} ), \sigma_\phi(\mathbf{h} ))
\end{align}
where $f_\phi$ is a neural network and often $\mu_\phi$ and $\sigma_\phi$ are simple linear transformations (but could also be neural networks). 

Since VAEs are parameterized by neural networks, we train them using gradient descent. Computing the gradient of the loss function with respect to $\theta$ is straightforward using automatic differentiation (we would only need to compute $\frac{\partial \log p_\theta(\mathbf{x_i} | \mathbf{z})}{\partial \theta}$ since the KL term is independent of $\theta$). However, computing the gradient with respect to $\phi$ is non-trivial since:
$$ \frac{\partial\ \text{ELBO}(\mathbf{x_i} | \theta, q_\phi)}{\partial \phi} = \frac{\partial\ \text{ELBO}(\mathbf{x_i} | \theta, q_\phi)}{\partial \mathbf{z}} \left[ \frac{\partial \mathbf{z}}{\partial \mu_{\mathbf{z}|\mathbf{x}}} \frac{\partial \mu_{\mathbf{z}|\mathbf{x}}}{\partial \phi} + \frac{\partial \mathbf{z}}{\partial \sigma_{\mathbf{z}|\mathbf{x}}} \frac{\partial \sigma_{\mathbf{z}|\mathbf{x}}}{\partial \phi}  \right]  $$
where we used chain rule and the notation from \myref{Figure}{fig:vae_diagram}.$\frac{\partial \mu_{\mathbf{z}|\mathbf{x}}}{\partial \phi}$ and $\frac{\partial \sigma_{\mathbf{z}|\mathbf{x}}}{\partial \phi}$ are trivial to differentiate using automatic differentiation, but how do we differentiate $\frac{\partial \mathbf{z}}{\partial \mu_{\mathbf{z}|\mathbf{x}}} $ and $\frac{\partial \mathbf{z}}{\partial \sigma_{\mathbf{z}|\mathbf{x}}} $? The difficulty arises from the fact that $\mathbf{z}$ is a random function of $\mu_{\mathbf{z}|\mathbf{x}}$ and $\sigma_{\mathbf{z}|\mathbf{x}}$. To ensure the sampling is differentiable, we use the \textit{reparameterization trick}: we can rewrite  sampling from $\mathbf{z} \sim \mathcal{N}(\mu_\phi(\mathbf{h}), \sigma_\phi(\mathbf{h}))$ (where the parameters are a part of the random function) as:
\begin{align}
    \mathbf{\epsilon} &\sim \mathcal{N}(0, 1)
\\ \mathbf{z} &= \mu_\phi(\mathbf{h}) + \sigma_\phi(\mathbf{h}) \cdot \mathbf{\epsilon}
\end{align}
The distribution is still the same, but now the parameters are not being directly used in the randomization. Thus, $\frac{\partial \mathbf{z}}{\partial \mu_{\mathbf{z}|\mathbf{x}}} = 1$ and $\frac{\partial \mathbf{z}}{\partial \sigma_{\mathbf{z}|\mathbf{x}}}  = \mathbf{\epsilon}$. The reparameterization trick can be generalized to any continuous distribution for which we know the inverse CDF function ($\mathcal{F}^{-1}$): 
\begin{align*}
    \mathbf{\epsilon} &\sim \mathcal{U}(0, 1)
\\ \mathbf{z} &= \mathcal{F}^{-1}(\mathbf{\epsilon})
\end{align*}
where $\mathcal{U}$ is a uniform distribution.

\subsection{Handling Missing Data}

Since the overall goal of our paper is to use VAEs for filling in missing data, we expect that our training data will also contain missing values. To handle missing data in probabilistic modeling, we must marginalize out (integrate out) the missing variables. 

For example, say our data is $\mathbf{x} = \mathbf{x}_\text{obs}\ ||\ \mathbf{x}_\text{miss}$ ($\mathbf{x}$ is the concatenation of a vector of observed data and a vector of missing data). Assume we are modeling the data using a deep latent variable model and that $\mathbf{x_i}\bigCI \mathbf{x_j} | \mathbf{z}$ (each variable is conditionally independent given $\mathbf{z}$). We then get:
\begin{align}
    \log p(\mathbf{x}_\text{obs} | \theta) &= \log \int p(\mathbf{x}_\text{obs}, \mathbf{x}_\text{miss} | \mathbf{z}, \theta) p(\mathbf{z}) \ d\mathbf{x}_\text{miss}\ d\mathbf{z}
\\ &= \log \int p(\mathbf{x}_\text{obs} | \mathbf{z}, \theta)\ p(\mathbf{x}_\text{miss} | \mathbf{z}, \theta)\ p(\mathbf{z}) \frac{q(\mathbf{z} | \mathbf{x}_\text{obs})}{q(\mathbf{z} | \mathbf{x}_\text{obs})} \ d\mathbf{x}_\text{miss}\  d\mathbf{z}
\\ &\geq \mathbb{E}_{\mathbf{z} \sim q(\mathbf{z} | \mathbf{x}_\text{obs})}\left[\log \int \frac{p(\mathbf{x}_\text{obs} | \mathbf{z}, \theta)\ p(\mathbf{x}_\text{miss} | \mathbf{z}, \theta)\ p(\mathbf{z})}{q(\mathbf{z} | \mathbf{x}_\text{obs})} \ d\mathbf{x}_\text{miss}\right]
\\ &= \mathbb{E}_{\mathbf{z} \sim q(\mathbf{z} | \mathbf{x}_\text{obs})}\left[\log \frac{p(\mathbf{x}_\text{obs} | \mathbf{z}, \theta)\ p(\mathbf{z})}{q(\mathbf{z} | \mathbf{x}_\text{obs})} \right]
\end{align}

Practically, in terms of implementation, the loss function with missing data is as simple as masking out the ``reconstruction'' terms for the missing values. While the change might seem trivial of an extension, the derivation above gives a simple but rigorous justification of the implementation.

\subsection{Reviewing the Definition of VAE}

Returning back to the definition we introduced earlier for VAEs, we can see that VAEs are based on two design decisions: 1) \textit{Using deep latent variable models} and 2) \textit{Training this model with variational inference}. Variational inference can be used to train other models such as a Gaussian Mixture Models or Latent Dirichlet Allocation \citep{blei2003lda} (neither of which relies on neural networks). Additionally, we could train DLVMs using other methods besides variational inference such as MCMC (Markov chain Monte Carlo), or even adversarial training. 
The view of a VAE as a regularized autoencoder is an overly specific description of the loss function, and we suggest that the fact that the loss function looks like a regularized autoencoder is more an outcome of the optimization process (using variational inference) and arguably is a coincidence (\myref{Equation}{eqn:normal_ll}). Finally, by centering our definition on a distributional definition instead of an autoencoder definition, we believe that the uncertainty-based component is better highlighted.

Realizing that there are many distributional components in a VAE, namely $q(\mathbf{z}| \mathbf{x})$, $p(\mathbf{z})$, and $p(\mathbf{x} | \mathbf{z})$, we could replace each distribution with something besides a Normal distribution. For example, we could use a more expressive family of distributions such as normalizing flows for $q(\mathbf{z}| \mathbf{x})$ \citep{Rezende2015NF}. Further, we could use a mixture distribution for $p(\mathbf{z})$ as was done by \citet{Jiang2017VADE} to allow the VAE to learn to cluster the data. We could use a multinomial distribution for $p(\mathbf{x} | \mathbf{z})$ if our data is categorical, or a Poisson if it is counts data \citep{nazabal2018mvae}, and even for continuous data, we could use a heavier tailed distribution such as a Logistic distribution (e.g., \citep{Vahdat2020NVAE}). Arguably, from the autoencoder viewpoint, these changes are not as obvious to make as they are from the perspective of distributional modeling.

\section{Why We Do Not Use Beta-VAE}\label{sec:beta_vae}

From the perspective of a VAE as a regularized autoencoder, a natural extension would be to add an additional hyperparameter to control the amount of regularization:
\begin{equation}\label{eqn:beta_vae}
 \mathcal{L}_{\beta\mhyphen \text{VAE}}(\mathbf{x};\beta, \theta, \phi,) = \mathbb{E}_{\mathbf{z} \sim q(\mathbf{z} | \mathbf{x})}\left[ \norm{\mathbf{x} - \mu_\theta(\mathbf{z} )}^2 \right] - \beta\ \text{KL}\left(q(\mathbf{z} | \mathbf{x_i}) || p(\mathbf{z})\right)
\end{equation}
Using this loss function is referred to as a Beta-VAE \citep{higgins2017betavae}, or $\beta$-VAE. While this is natural in the autoencoder perspective and has been used many times in the financial literature for imputation, what is the interpretation in the distributional perspective?

We repeat the VAE loss for the specific generative process (\myref{Equation}{eqn:vanilla_gen}) here:
\begin{equation}\label{eqn:sigma_vae}
 \mathcal{L}_{\sigma\mhyphen \text{VAE}}(x_i;\sigma_x),  \theta, \phi) = \mathbb{E}_{z \sim q(z | x_i)}\left[\frac{1}{2}\norm{\frac{\mathbf{x_i} - \mu_\theta(\mathbf{z})}{\sigma_x}}^2_2  + \log \left(\sigma_x d \sqrt{2 \pi}\right) \right] - \text{KL}\left(q(\mathbf{z} | \mathbf{x_i}) || p(\mathbf{z})\right)
\end{equation}
By comparing and contrasting these two loss functions, we will be able to better understand why $\beta$-VAE has been useful, and, more importantly, how we generalize it to be more effective in practice.

First, let's assume that $\sigma_x$ is fixed and not a learnable parameter. To compare $\mathcal{L}_{\beta-\text{VAE}}$ and $\mathcal{L}_{\sigma-\text{VAE}}$, we remind ourselves that it isn't crucial to compare the value of the two loss functions, but instead to compare the minima of the two:
\begin{align}
 \argmin_{\theta, \phi} \mathcal{L}_{\sigma\mhyphen\text{VAE}}(\mathbf{x_i}; \sigma_x, \theta, \phi) &= \argmin_{\theta, \phi} \mathbb{E}_{z \sim q(z | x_i)}\left[\frac{1}{2}\norm{\frac{\mathbf{x_i} - \mu_\theta(\mathbf{z})}{\sigma_x}}^2_2  + \log \left(\sigma_x d \sqrt{2 \pi}\right) \right] - \text{KL}\left(q(\mathbf{z} | \mathbf{x_i}) || p(\mathbf{z})\right)
\\ &= \argmin_{\theta, \phi} \mathbb{E}_{z \sim q(z | x_i)}\left[\frac{1}{2 \sigma_x^2 }\norm{\mathbf{x_i} - \mu_\theta(\mathbf{z})}^2 \right] - \text{KL}\left(q(\mathbf{z} | \mathbf{x_i}) || p(\mathbf{z})\right)
\\ &= \argmin_{\theta, \phi} \mathbb{E}_{z \sim q(z | x_i)}\left[\norm{\mathbf{x_i} - \mu_\theta(\mathbf{z})}^2 \right] - 2 \sigma_x^2 \text{KL}\left(q(\mathbf{z} | \mathbf{x_i}) || p(\mathbf{z})\right)
\\ &= \argmin_{\theta, \phi} \mathcal{L}_{\beta\mhyphen \text{VAE}}(\mathbf{x_i}; 2 \sigma_x^2 , \theta, \phi)
\end{align}
In other words, optimizing a $\sigma\mhyphen\text{VAE}$ with a fixed $\sigma_x$ is equivalent to optimizing a $\beta$-VAE where $\beta = 2 \sigma_x^2$. Thus, from the distributional perspective, $\beta$ simply controls the variance of the Normal distribution on the decoder side.

While this might seem simply to be an interesting mathematical connection, there is one major reason to use $\sigma\mhyphen\text{VAE}$ instead of $\beta$-VAE: we can learn $\sigma$ instead of tune it. If $\sigma_x$ is not fixed, there is an additional term that $\beta$-VAE does not have: $\log \left(\sigma_x d \sqrt{2 \pi}\right)$. This ``correction term'' allows us to optimize $\sigma_x$ instead of treating it as a hyperparameter that must be tuned.

We run a simple experiment to highlight this point. We train four models:
\begin{itemize}
    \item \textit{Model 1}: $\beta$-VAE with $\beta=1$
    \item \textit{Model 2}: $\beta$-VAE with $\beta=0.1$
    \item \textit{Model 3}: $\beta$-VAE with learnable $\beta$
    \item \textit{Model 4}: $\sigma$-VAE with learnable $\sigma$
\end{itemize}
We show the (synthetic) data we trained our VAEs on in \myref{Figure}{fig:two_dimensional_gauss_data}.
To ensure no underfitting, we trained all models for 100K steps. Additional details on architecture and training methodology can be found in \myref{Appendix}{sec:exp_detail_beta_vae}.

\begin{figure}[!bt]
\begin{minipage}{\linewidth}
 \centerline{\includegraphics[width=0.3\linewidth]{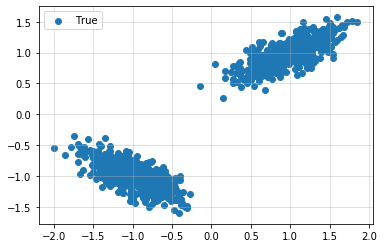}}
    \caption{Mixture of two Gaussians.}
\label{fig:two_dimensional_gauss_data}
\end{minipage}
\begin{minipage}{\linewidth}
    \begin{subfigure}{0.24\linewidth}
    \centerline{\includegraphics[width=\linewidth]{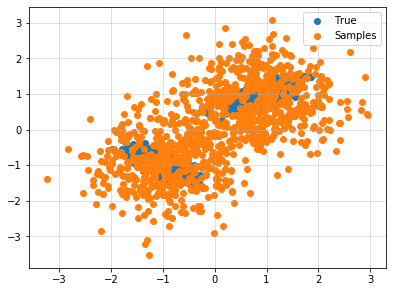}}
    \caption{$\beta=1$}
    \end{subfigure}
    \begin{subfigure}{0.24\linewidth}
    \centerline{\includegraphics[width=\linewidth]{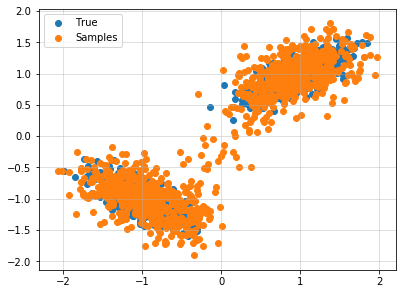}}
    \caption{$\beta=0.1$}
    \end{subfigure}
    \begin{subfigure}{0.24\linewidth}
    \centerline{\includegraphics[width=\linewidth]{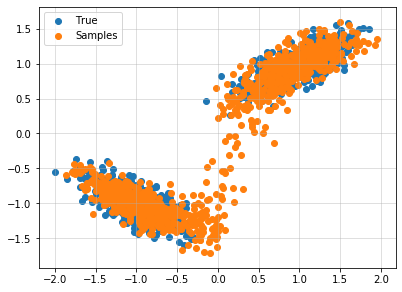}}
    \caption{Learnable $\beta$}
    \end{subfigure}
    \begin{subfigure}{0.24\linewidth}
    \centerline{\includegraphics[width=\linewidth]{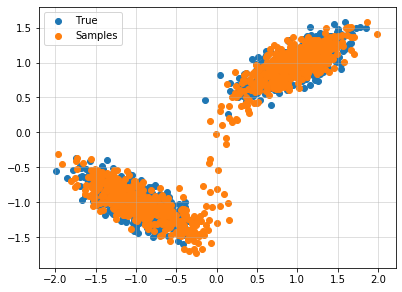}}
    \caption{$\sigma$-VAE}
    \end{subfigure}
    \caption{Comparison of generations by different VAEs.}\label{fig:2d_generations}
\end{minipage}

\end{figure}

To create the generations for the $\beta$-VAE that we see in \myref{Figure}{fig:2d_generations}, we modify the variance $\sigma^2_x$ to be $\sigma_x^2 = \frac{\beta}{2}$. We can see that the quality of generations with $\beta=1$ is low as a result of not training the output variance ($\sigma_x$). We see that lowering $\beta$ helps improve the generations. However, for the learnable $\beta$, the final $\beta$ was $1e{-4}$, which is to be expected as it is optimal to remove the KL-term when minimizing the $\beta$-VAE loss. From a practical standpoint, this is suboptimal as the intention of $\beta$ is to ``regularize'' the latent space. On the other hand, for the $\sigma$-VAE, the quality of the generations is good (and will be improved in a later section (\myref{Section}{sec:iwae})) and has a final learned $\sigma_x=0.127$. Thus, while good generations might be attainable by a $\beta$-VAE, finding the optimal $\beta$ requires training multiple models, and tuning $\beta$ whereas \textit{$\sigma$-VAE requires training only once}. We show in the next section (\myref{Section}{sec:hetero}) that heteroscedastic noise is easy to account for in a $\sigma$-VAE, but is impractical for $\beta$-VAE due to the massive increase in hyperparameters.

\section{Improving Training Methodology of VAEs}

\subsection{Heteroscedasticity}\label{sec:hetero}

Having realized that we can simply use $\sigma\mhyphen VAE$ to remove the need to hyperparameter tune $\beta$ in $\beta\mhyphen VAE$ (\myref{Section}{sec:beta_vae}), we can easily extend the VAE to allow for each feature to have a different amount of noise. We can easily replace the scalar $\sigma_x$ with a vector $\mathbf{\sigma_x} \in \RR^p$ and optimize the amount of noise for each feature jointly with all the other parameters ($\theta$ and $\phi$):
\begin{equation}\label{eqn:sigma_vec_vae}
 \mathcal{L}_{\mathbf{\vec{\sigma}}\mhyphen \text{VAE}}(\mathbf{x_i};\sigma_x, \theta, \phi) = \mathbb{E}_{z \sim q(z | x_i)}\left[\frac{1}{2}\norm{\frac{\mathbf{x_i} - \mu_\theta(\mathbf{z})}{\mathbf{\sigma_x}}}^2_2  +\sum_{i=1}^{p} \log \left( \sqrt{2 \pi} \sigma_{x,i} \right) \right] - \text{KL}\left(q(\mathbf{z} | \mathbf{x_i}) || p(\mathbf{z})\right)
\end{equation}
If this were a $\beta$-VAE, this would introduce a hyperparameter per input dimension (feature), i.e., $p$ hyperparameters; here, with the distributional perspective of VAEs, we introduce \textit{no new hyperparameters}.

Further, we can make $\mathbf{\sigma_x}$ a function of $z$:
\begin{equation}\label{eqn:sigma_fn_vae}
 \mathcal{L}_{{{\Sigma}}\mhyphen \text{VAE}}(\mathbf{x_i},  \theta, \phi) = \mathbb{E}_{z \sim q(z | x_i)}\left[\frac{1}{2}\norm{\frac{(\mathbf{x_i} - \mu_\theta(\mathbf{z}))}{\sigma_\theta(z)}}^2_2  +\sum_{i=1}^{p} \log \left( \sqrt{2 \pi} (\sigma_\theta(z))_i \right) \right] - \text{KL}\left(q(\mathbf{z} | \mathbf{x_i}) || p(\mathbf{z})\right)
\end{equation}

To test the effect of this, similar to \myref{Section}{sec:beta_vae}, we generate a mixture of two Gaussians; however, we set the variances of the two modes to be different (\myref{Figure}{fig:two_dimensional_gauss_data2}). In \myref{Figure}{fig:2d_generations2}, we compare training with a single $\sigma_x$ versus a function $\mathbf{\sigma_\theta}$. While it might not be clear from the generations the difference in performance, the loss curve shows clearly the $\Sigma$-VAE outperforming. Additional details on architecture and training methodology can be found in \myref{Appendix}{sec:exp_detail_hetero_vae}.

\begin{figure}[!bt]
\begin{minipage}{\linewidth}
 \centerline{\includegraphics[width=0.3\linewidth]{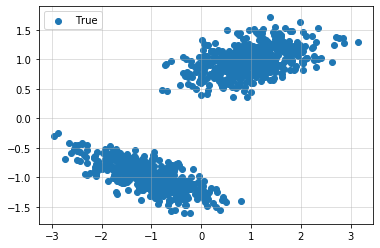}}
    \caption{Mixture of two Gaussians.}
\label{fig:two_dimensional_gauss_data2}
\end{minipage}
\begin{minipage}{\linewidth}
    \centerline{\includegraphics[width=\linewidth]{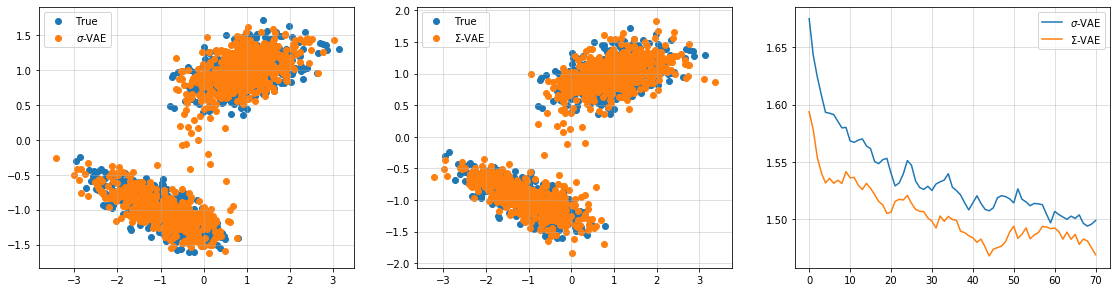}}
    \begin{minipage}[t]{.33\linewidth}
    \centering
        \vspace*{-2mm}
    \subcaption{$\sigma$-VAE}
    \end{minipage}%
    \begin{minipage}[t]{.33\linewidth}
    \centering
        \vspace*{-2mm}
    \subcaption{$\Sigma$-VAE}
    \end{minipage}
    \begin{minipage}[t]{.33\linewidth}
    \centering
        \vspace*{-2mm}
    \subcaption{Loss}
        \end{minipage}
    \caption{Comparison of generations and loss curves of two different VAEs.}\label{fig:2d_generations2}
\end{minipage}

\end{figure}

\subsection{Importance Weighted Autoencoder (IWAE)}\label{sec:iwae}

A simple trick to help close the gap in the lower bound is to use importance weighted autoencoders:
\begin{equation}\label{eqn:iwae_loss}
\log p(\mathbf{x_i} | \theta) \geq \mathbb{E}_{\mathbf{z_1}, \dots, \mathbf{z_k} \sim q(\mathbf{z} | \mathbf{x_i})}\left[ \log \sum_j \frac{1}{k} \frac{p(\mathbf{x_i} | \mathbf{z_j}, \theta) p(\mathbf{z_j})}{q(\mathbf{z_j} | \mathbf{x_i})}\right] =  -\mathcal{L}_{\text{IWAE}, k}
\end{equation}

As shown by \citet{burda2015iwae}, with some abuse of notation, we have
\begin{align}
\log p(\mathbf{x_i} | \theta) &=  -\mathcal{L}_{\text{IWAE}, \infty} 
\\ &\geq \mathcal{L}_{\text{IWAE}, k + 1} \geq -\mathcal{L}_{\text{IWAE}, k} 
\\ &\geq -\mathcal{L}_{\text{IWAE}, 1} = -\mathcal{L}_{\text{VAE}}
\end{align}
which implies using a larger $k$ reduces the bias of the estimator.

Note:
\begin{align}
    \log p(\mathbf{x_i} | \theta) &\geq  -\mathcal{L}_{\text{VAE}}
\\ &= \mathbb{E}_{\mathbf{z} \sim q(\mathbf{z} | \mathbf{x_i})}\left[ \log  \frac{p(\mathbf{x_i} | \mathbf{z}, \theta) p(\mathbf{z})}{q(\mathbf{z} | \mathbf{x_i})}\right]  
\\ &= \mathbb{E}_{\mathbf{z_1}, \dots, \mathbf{z_m} \sim q(\mathbf{z} | \mathbf{x_i})}\left[ \sum_j \frac{1}{m} \log  \frac{p(\mathbf{x_i} | \mathbf{z_j}, \theta) p(\mathbf{z_j})}{q(\mathbf{z_j} | \mathbf{x_i})}\right]
\end{align}
where, as opposed to the IWAE loss (\myref{Equation}{eqn:iwae_loss}), sampling $m$ times from $q(\mathbf{z} | \mathbf{x_i})$ does not reduce the bias but instead reduces the variance of the estimator of the VAE loss.

Using the synthetic data shown in \myref{Figure}{fig:two_dimensional_gauss_data}, we show in  \myref{Figure}{fig:iwae_2d_example} the effect of using importance weighted autoencoders on bimodal data. Additional details on architecture and training methodology can be found in \myref{Appendix}{sec:exp_detail_beta_vae}.
While the VAE struggles to fully capture the bimodality of the data, the IWAE is better able to handle this, leading to higher quality generations.

\begin{figure}[!bt]
\begin{minipage}{\linewidth}
    
    \begin{subfigure}{0.45\linewidth}
    \centerline{\includegraphics[width=\linewidth]{figures/generations_k=1_use_resid_block=True_beta_vae=None_learnable_beta=False.png}}
    \caption{$\sigma$-VAE}
    \end{subfigure}\hfill
    \begin{subfigure}{0.45\linewidth}
    \centerline{\includegraphics[width=\linewidth]{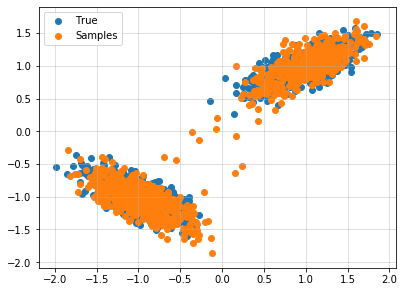}}
    \caption{$\sigma$-IWAE}
    \end{subfigure}
    \caption{Comparison of generations by  VAEs vs. IWAEs}\label{fig:iwae_2d_example}
\end{minipage}
\end{figure}

Note, while our experiments in \myref{Section}{sec:perf_iwae} show that training IWAEs does not help for FX volatility imputation, we do still utilize the IWAE loss for evaluation since we want an accurate estimate of $\log p(\mathbf{x_i} | \theta)$ for model comparison.

\subsection{Residual Networks}\label{sec:resid}

An additional adjustment we make, and the most important according to our experiments (\myref{Section}{sec:perf_resid}), is changing the architecture to residual networks. Normally, for tabular data, a feed forward network is used. However, with the continued success of residual networks in the context of convolutions \citep{he2016resnet,zagoruyko2017wideresnet}, transformers \citep{vaswani2017attention}, and normalizing flows (i.e., residual flows, \citet{Chen2019ResidualFlows}), we use residual networks as the backbone of our model, similar to the MLP blocks used in transformers (\myref{Figure}{fig:mlp_block}). Preliminary experiments showed better convergence when utilizing Swish ($=x * \text{Sigmoid}(x)$) \citep{ramachandran2018swish} over GELU and ReLU. We incorporate dropout after the activation layer and the second linear layer in the residual branch, similar to Transformers. A similar architecture has been previously introduced by \citet{Gorishniy2021TabRes}. However, \citet{Gorishniy2021TabRes} used Batch Normalization \citep{ioffe2015batchnorm}
instead of Layer Normalization \citep{Ba2016LayerNorm} and used ReLU instead of Swish.

\begin{figure}[!bt]
\begin{minipage}{\linewidth}
\begin{center}
\begin{tikzpicture}[
  every neuron/.style={
    circle,
    minimum size=0.3cm,
    very thick
  },
  every data/.style={
    rectangle,
    minimum size=0.4cm,
    thick
  },
]

  \node [align=center,every neuron/.try, data 1/.try, minimum width=0.3cm] (input-1)  at ($ (2.5, 0) $) {$\mathbf{x}\in\RR^{d}$};

  \node [align=center,every data/.try, data 1/.try, minimum height=2.0cm,draw] (input-2)  at ($ (input-1) + (3.0, 0) $) {$\sigma$};
  \node [align=center,every data/.try, data 1/.try, minimum height=2.0cm] (input-3)  at ($ (input-2) + (3.0, 0) $) {$\mathbf{y} \in \RR^{o}$};



  \draw [black,solid,->] (input-1) -- ($(input-2.west)$)node[midway,above] {$W_1 \in \RR^{h \times d}$};
  \draw [black,solid,->] (input-2) -- (input-3) node[midway,above] {$W_2 \in \RR^{h \times o}$};

\end{tikzpicture}
\end{center}
\caption{Diagram showing a traditional MLP block. $\sigma$ denotes a non-linear activation.}\label{fig:mlp_block}
\end{minipage}
\begin{minipage}{\linewidth}
\begin{center}
\begin{tikzpicture}[
  every neuron/.style={
    circle,
    minimum size=0.3cm,
    very thick
  },
  every data/.style={
    rectangle,
    minimum size=0.4cm,
    thick
  },
]

  \node [align=center,every neuron/.try, data 1/.try, minimum width=0.3cm] (input-1)  at ($ (2.5, 0) $) {$\mathbf{x}$};

  \node [align=center,every data/.try, data 1/.try, minimum height=2.0cm,draw] (input-2)  at ($ (input-1) + (1.5, 0) $) {$\sigma$};
  \node [align=center,every data/.try, data 1/.try, minimum height=2.0cm,draw] (input-3)  at ($ (input-2) + (1.5, 0) $) {$\sigma$};
  \node [align=center,every data/.try, data 1/.try, minimum height=2.0cm] (input-4)  at ($ (input-3) + (1.5, 1.0) $) {$\mu_{z|x}$};
  \node [align=center,every data/.try, data 1/.try, minimum height=2.0cm] (input-5)  at ($ (input-3) + (1.5, -1.0) $) {$\sigma_{z|x}$};
  
  \node [align=center,every data/.try, data 1/.try, minimum height=2.0cm] (input-6)  at ($ (input-5) + (1.5, 1.0) $) {$z$};
  \node [align=center,every data/.try, data 1/.try, minimum height=2.0cm,draw] (input-7)  at ($ (input-6) + (1.5, 0) $) {$\sigma$};
  \node [align=center,every data/.try, data 1/.try, minimum height=2.0cm,draw] (input-8)  at ($ (input-7) + (1.5, 0) $) {$\sigma$};
\node [align=center,every data/.try, data 1/.try, minimum height=2.0cm] (input-9)  at ($ (input-8) + (1.5, 1.0) $) {$\mu_{x|z}$};
  \node [align=center,every data/.try, data 1/.try, minimum height=2.0cm] (input-10)  at ($ (input-8) + (1.5, -1.0) $) {$\sigma_{x|z}$};



  \draw [black,solid,->] (input-1) -- ($(input-2.west)$)node[midway,above] {$W_1$};
  \draw [black,solid,->] (input-2) -- (input-3) node[midway,above] {$W_2$};
  \draw [black,solid,->] (input-3) -- (input-4) node[midway,above, yshift=0.1cm] {$W_{3,1}$};
  \draw [black,solid,->] (input-3) -- (input-5) node[midway,above] {$W_{3,2}$};
  \draw [black,solid,->] (input-4) -- (input-6) ;

  \draw [black,solid,->] (input-5) -- (input-6) node[midway,above, yshift=0.1cm,xshift=-0.3cm] {Sample};

  \draw [black,solid,->] (input-6) -- ($(input-7.west)$)node[midway,above] {$W_4$};
  \draw [black,solid,->] (input-7) -- (input-8) node[midway,above] {$W_5$};
  \draw [black,solid,->] (input-8) -- (input-9) node[midway,above, yshift=0.1cm] {$W_{6,1}$};
  \draw [black,solid,->] (input-8) -- (input-10) node[midway,above] {$W_{6,2}$};

\end{tikzpicture}
\end{center}
\caption{Diagram showing a VAE comprised of traditional MLP blocks. }\label{fig:mlp_vae}
\end{minipage}
\end{figure}

\begin{figure}[!bt]
\begin{minipage}{\linewidth}
\begin{center}
\begin{tikzpicture}[
  every neuron/.style={
    circle,
    minimum size=0.3cm,
    very thick
  },
  every data/.style={
    rectangle,
    minimum size=0.4cm,
    thick
  },
]

  \node [align=center,every neuron/.try, data 1/.try, minimum width=0.3cm] (input-1)  at ($ (2.5, 0) $) {};

  \node [align=center,every neuron/.try, data 1/.try] (input-0)  at ($ (input-1) - (2.5, 0) $) {$\mathbf{x}\in \RR^{d}$};

  \node [align=center,every data/.try, data 1/.try, minimum height=2.0cm,draw] (input-2)  at ($ (input-1) + (1.5, -2) $) {$\sigma$};


  \node [align=center,every neuron/.try, data 1/.try, minimum width=0.3cm,draw] (plus)  at ($ (input-2) + (1.5, 2) $) {$\mathbf{+}$};
  \node [align=center,every neuron/.try, data 1/.try, minimum width=0.3cm] (output)  at ($ (plus) + (2.5, 0) $) {$y \in \RR^{o}$};

  \draw [black,solid,->] (input-0) -- ($(input-1.west)$) node[midway,above] {$W_1\in\RR^{h\times d}$};
  \draw [black,solid,->] (input-1) -- ($(input-2.west)$) node[midway,left,align=right] {\text{Layer Norm }\\ \\ $ W_2\in\RR^{h\times h}$};
  \draw [black,solid,->] (input-1) -- (plus);
  \draw [black,solid,->] ($(input-2.east)$) -- (plus) node[midway,right] {$W_3\in\RR^{h\times h}$};
  \draw [black,solid,->] (plus) -- ($(output.west)$)node[midway,above] {$W_4\in\RR^{d\times o}$};

\end{tikzpicture}
\end{center}
\caption{Diagram showing a residual block. In our model, we use square matrices for $W_1$ and $W_2$, and we use Swish for the activation $\sigma$.}\label{fig:resid_block}
\end{minipage}
\begin{minipage}{\linewidth}
\begin{center}
\scalebox{.8}{
\begin{tikzpicture}[
  every neuron/.style={
    circle,
    minimum size=0.3cm,
    very thick
  },
  every data/.style={
    rectangle,
    minimum size=0.4cm,
    thick
  },
]

  \node [align=center,every neuron/.try, data 1/.try, minimum width=0.3cm] (input-1)  at ($ (1.5, 0) $) {};

  \node [align=center,every neuron/.try, data 1/.try] (input-0)  at ($ (input-1) - (1.0, 0) $) {$\mathbf{x}$};

  \node [align=center,every data/.try, data 1/.try, minimum height=2.0cm,draw] (input-2)  at ($ (input-1) + (1.5, -1.5) $) {$\sigma$};
  \node [align=center,every neuron/.try, data 1/.try, minimum width=0.3cm,draw] (plus-1)  at ($ (input-2) + (1.5, 1.5) $) {$\mathbf{+}$};

  \node [align=center,every data/.try, data 1/.try, minimum height=2.0cm,draw] (input-3)  at ($ (plus-1) + (1.5, -1.5) $) {$\sigma$};
  \node [align=center,every neuron/.try, data 1/.try, minimum width=0.3cm,draw] (plus-2)  at ($ (input-3) + (1.5, 1.5) $) {$\mathbf{+}$};

  \node [align=center,every data/.try, data 1/.try, minimum height=2.0cm] (input-4)  at ($ (plus-2) + (1.5, 1.0) $) {$\mu_{z|x}$};
  \node [align=center,every data/.try, data 1/.try, minimum height=2.0cm] (input-5)  at ($ (plus-2) + (1.5, -1.0) $) {$\sigma_{z|x}$};
  \node [align=center,every data/.try, data 1/.try, minimum height=2.0cm] (input-6)  at ($ (input-5) + (1.5, 1.0) $) {$z$};

  \node [align=center,every neuron/.try, data 1/.try, minimum width=0.3cm] (input-7)  at ($(input-6) + (1.0, 0) $) {};

  \node [align=center,every data/.try, data 1/.try, minimum height=2.0cm,draw] (input-8)  at ($ (input-7) + (1.5, -1.5) $) {$\sigma$};
  \node [align=center,every neuron/.try, data 1/.try, minimum width=0.3cm,draw] (plus-3)  at ($ (input-8) + (1.5, 1.5) $) {$\mathbf{+}$};

  \node [align=center,every data/.try, data 1/.try, minimum height=2.0cm,draw] (input-9)  at ($ (plus-3) + (1.5, -1.5) $) {$\sigma$};
  \node [align=center,every neuron/.try, data 1/.try, minimum width=0.3cm,draw] (plus-4)  at ($ (input-9) + (1.5, 1.5) $) {$\mathbf{+}$};
  \node [align=center,every data/.try, data 1/.try, minimum height=2.0cm] (input-10)  at ($ (plus-4) + (1.5, 1.0) $) {$\mu_{x|z}$};
  \node [align=center,every data/.try, data 1/.try, minimum height=2.0cm] (input-11)  at ($ (plus-4) + (1.5, -1.0) $) {$\sigma_{x|z}$};


  \draw [black,solid,->] (input-0) -- ($(input-1.west)$) node[midway,above] {$W_1$};
  \draw [black,solid,->] (input-1) -- ($(input-2.west)$) node[midway,align =right, left] {LN \\ $W_2$};
  \draw [black,solid,->] (input-1) -- (plus-1);
  \draw [black,solid,->] (plus-1) -- (plus-2);
  \draw [black,solid,->] ($(input-2.east)$) -- (plus-1) node[midway,right] {$W_3$};
  \draw [black,solid,->] (plus-1) -- ($(input-3.west)$) node[midway,align =left, left] {LN \\ $W_4$};
  \draw [black,solid,->] (input-3) -- ($(plus-2.west)$) node[midway,right] {$W_5$};

  \draw [black,solid,->] (plus-2) -- (input-4) node[midway,above, yshift=0.1cm] {$W_{6,1}$};
  \draw [black,solid,->] (plus-2) -- (input-5) node[midway,above] {$W_{6,2}$};
  \draw [black,solid,->] (input-4) -- (input-6) ;

  \draw [black,solid,->] (input-5) -- (input-6) node[midway,above, yshift=0.1cm,xshift=-0.3cm] {Sample};

  \draw [black,solid,->] (input-6) -- (input-7) node[midway,above] {$W_7$};
  \draw [black,solid,->] (input-7) -- (input-8) node[midway,align =right, left] {LN \\ $W_8$};
  \draw [black,solid,->] (input-7) -- (plus-3) ;
  \draw [black,solid,->] (input-8) -- (plus-3) node[midway,right] {$W_9$};
  \draw [black,solid,->] (plus-3) -- (plus-4) ;
  \draw [black,solid,->] (plus-3) -- (input-9) node[midway,align =right, left] {LN \\ $W_{10}$};
  \draw [black,solid,->] (input-9) -- (plus-4) node[midway,right] {$W_{11}$};
  \draw [black,solid,->] (plus-4) -- (input-10) node[midway,above, yshift=0.1cm] {$W_{12,1}$};
  \draw [black,solid,->] (plus-4) -- (input-11) node[midway,below] {$W_{12,2}$};


\end{tikzpicture}
}
\end{center}
\caption{Diagram showing a VAE comprised of residual blocks. We use ``LN'' to denote Layer Normalization.}\label{fig:resid_vae}
\end{minipage}
\end{figure}

To make the differences clearer, we show the conventional approach to MLPs in \myref{Figure}{fig:mlp_block} and \myref{Figure}{fig:mlp_vae} and show the approach using residual networks in  in \myref{Figure}{fig:resid_block} and \myref{Figure}{fig:resid_vae}. Mathematical notation describing the VAE comprised of residual networks can be found in \myref{Appendix}{sec:vae_resid_arch}.

To compare the two architectures, we fit them to a mixture of eight Gaussians (\myref{Figure}{fig:eight_gauss_data}).
In \myref{Figure}{fig:eight_gauss_gen}, we see that the residual network is able to easily learn this distribution while the normal feedforward struggles. We see this both from the quality of samples where the residual network is better able to capture the multimodality (\myref{Figure}{sec:resid_comparison_mlp} and \myref{Figure}{sec:resid_comparison_resid}) as well as in the gap in loss curves (\myref{Figure}{sec:resid_comparison_loss}). Additional details on architecture and training methodology can be found in \myref{Appendix}{sec:exp_detail_resid_vae}.

\begin{figure}[!bt]
\begin{minipage}{\linewidth}
 \centerline{\includegraphics[width=0.3\linewidth]{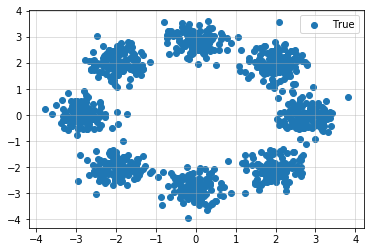}}
    \caption{Mixture of eight Gaussians.}
\label{fig:eight_gauss_data}
\end{minipage}
\begin{minipage}{\linewidth}
    \centerline{\includegraphics[width=\linewidth]{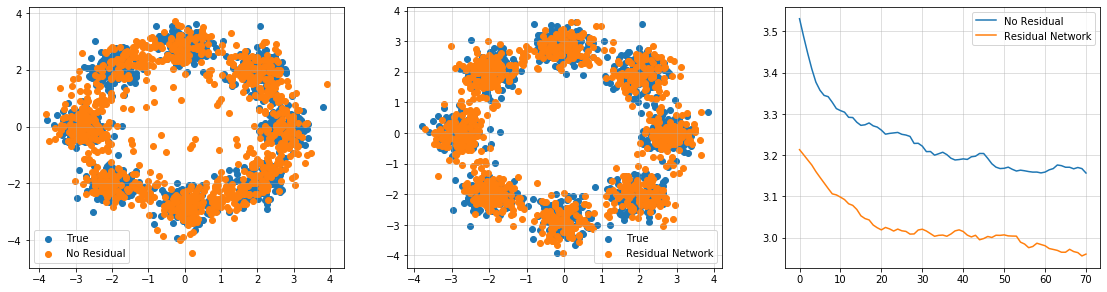}}
    \begin{minipage}[t]{.33\linewidth}
    \centering
        \vspace*{-2mm}
    \subcaption{MLP VAE}\label{sec:resid_comparison_mlp}
    \end{minipage}%
    \begin{minipage}[t]{.33\linewidth}
    \centering
        \vspace*{-2mm}
    \subcaption{Residual VAE}\label{sec:resid_comparison_resid}
    \end{minipage}
    \begin{minipage}[t]{.33\linewidth}
    \centering
        \vspace*{-2mm}
    \subcaption{Loss}\label{sec:resid_comparison_loss}
        \end{minipage}
    \caption{Comparison of generations and loss curves with and without using residual blocks.}\label{fig:eight_gauss_gen}
\end{minipage}
\end{figure}

\subsubsection{Intuition of Efficacy of Residual Networks}

To build an intuition on the residual architecture, looking at \myref{Figure}{fig:resid_vae}, we can see, from a gradient perspective, there is ``direct path'' from the reconstruction loss to $\mu_{\mathbf{z} | \mathbf{x}}$ as opposed to \myref{Figure}{fig:mlp_vae}. We analyze the impact of this architectural detail by analyzing:
$$ \norm{\frac{1}{N} \sum_{i=1}^{N} \frac{\partial \mathbb{E}_{\mathbf{z} \sim q(\mathbf{z} | \mathbf{x_i})}\left[\log p(\mathbf{x_i} | \mathbf{z}) \right]}{\partial \mu_{\mathbf{z} | \mathbf{x}}}}_2 $$
which we will refer to as the gradient norm of the reconstruction term, and:
$$\qquad \norm{\frac{1}{N} \sum_{i=1}^{N} \frac{\partial \text{KL}\left(q(\mathbf{z} | \mathbf{x_i})\ ||\ p(\mathbf{z})\right)}{\partial \mu_{\mathbf{z}| \mathbf{x}}}}_2 $$
which we will refer to as the gradient norm of the KL term. In \myref{Figure}{fig:grad_norm_resid}, we plot these two quantities (averaged across 500 batches) as a function of training steps for a VAE with and without residual connections. 
We find that:
\begin{enumerate}
    \item The gradient norm of the reconstruction term starts small for VAEs without residual connections
    \item The gradient norm of the reconstruction term on average is larger with residual connections than without. Further, the gradient norm of the KL term is comparable. 
\end{enumerate}
The second observation is interesting since the original intention of $\beta$-VAE was to reduce the regularization from the KL term which is equivalent to reducing the gradient norm of the KL term. By using residual connections, we get this behavior without having to modify the loss function, retaining the probabilistic interpretation of VAEs.

\begin{figure}[!bt]
\begin{minipage}{\linewidth}
 \centerline{\includegraphics[width=0.8\linewidth]{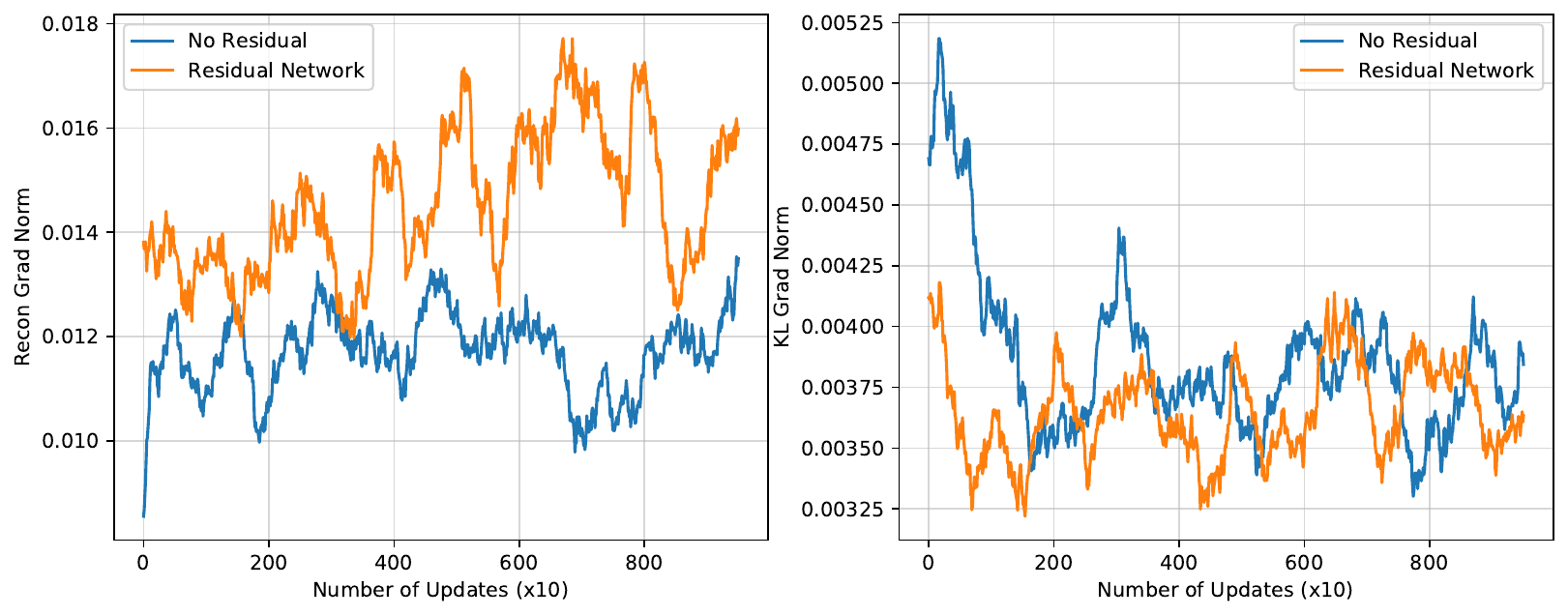}}
    \caption{Comparison of gradient norm of reconstruction term and KL term for VAEs with and without residual connections.}
\label{fig:grad_norm_resid}
\end{minipage}

\end{figure}

Further, if we were to remove all the residual paths (the paths from the Layer Norm to the layer before the elementwise addition), we would get a linear model, and a linear VAE has connections to probabilistic PCA \citep{Ghojogh2021FactorPCA} and factor analysis. Given the efficacy of PCA in finance, we intuit this connection to be a source of the performance we see in \myref{Section}{sec:perf_resid}.

\section{Distributional Imputation}\label{sec:imputation}

Since we are focused on uncertainties and probabilities, we view imputation as querying $p(\mathbf{x_m} | \mathbf{x_o})$ where $\mathbf{x_m}$ and $\mathbf{x_o}$ is a partitioning of $\mathbf{x}$ where $\mathbf{x_o}$ is the subset of features observed and $\mathbf{x_m}$  is the subset of features missing. We focus on $p(\mathbf{x_m} | \mathbf{x_o})$ generically as we can view point estimates as the mean $\EE{p(\mathbf{x_m} | \mathbf{x_o})}$ and uncertainty estimates as the variance $\VV{p(\mathbf{x_m} | \mathbf{x_o})}$.

\subsection{Algorithm and Derivation}\label{sec:dist_imp_algo}

The question then is, how do we sample from $p(\mathbf{x_m} | \mathbf{x_o})$? To begin, we first use the law of total probability:
$$p(\mathbf{x_m} | \mathbf{x_o}) =\int p(\mathbf{x_m} | \mathbf{x_o}, \mathbf{z})\ p(\mathbf{z}| \mathbf{x_o})\ d\mathbf{z} $$
Algorithmically, we can read the integral as:
\begin{enumerate}
    \item Sample $\mathbf{z}$ from $p(\mathbf{z}| \mathbf{x_o})$
    \item Sample $\mathbf{x_m} $ from $p(\mathbf{x_m} |\mathbf{x_o}, \mathbf{z})$
\end{enumerate}

Note, $p(\mathbf{x_m} |\mathbf{x_o}, \mathbf{z}) = p(\mathbf{x_m} |\mathbf{z})$ by construction of our VAE since, conditional on $\mathbf{z}$, $\mathbf{x}$ is distributed as $p$-independent Gaussians.

To sample from $p(\mathbf{z}| \mathbf{x_o})$, we see using Bayes rule that $p(\mathbf{z}| \mathbf{x_o}) \propto p(\mathbf{x_o}| \mathbf{z}) p(\mathbf{z})$, or algorithmically:
\begin{enumerate}
    \item Sample $\mathbf{z_i}$ from $p(\mathbf{z})$ $n$ times
    \item Reweigh each $\mathbf{z_i}$ by $\frac{p(\mathbf{x_o}| \mathbf{z_i})}{\sum_j p(\mathbf{x_o}| \mathbf{z_j})}$
\end{enumerate}
However, since $p(\mathbf{z}| \mathbf{x_o})$ is the posterior and the encoder $q(\mathbf{z} | \mathbf{x_o})$ is meant to be an approximation of the posterior, we can rewrite:
$$p(\mathbf{z}| \mathbf{x_o}) \propto p(\mathbf{x_o}| \mathbf{z}) p(\mathbf{z}) \frac{q(\mathbf{z} | \mathbf{x_o})}{q(\mathbf{z} | \mathbf{x_o})}$$
\begin{enumerate}
    \item Sample $z_i$ from $q(\mathbf{z} | \mathbf{x_o})$ $n$ times
    \item Reweigh each $z_i$ by $\frac{p(\mathbf{x_o}| \mathbf{z_i}) p(\mathbf{z_i})}{q(\mathbf{z_i} | \mathbf{x_o})}/\sum_j \frac{p(\mathbf{x_o}| \mathbf{z_j}) p(\mathbf{z_j})}{q(\mathbf{z_j} | \mathbf{x_o})} = w_i$
\end{enumerate}

Plugging all this into the original algorithm, we get
\begin{enumerate}
    \item Sample $\mathbf{z_i}$ from $q(\mathbf{z} | \mathbf{x_o})$ $n$ times
    \item Reweigh each $\mathbf{z_i}$ by $\frac{p(\mathbf{x_o}| \mathbf{z_i}) p(\mathbf{z_i})}{q(\mathbf{z_i} | \mathbf{x_o})}/\sum_j \frac{p(\mathbf{x_o}| \mathbf{z_j}) p(\mathbf{z_j})}{q(\mathbf{z_j} | \mathbf{x_o})} = w_i$
    \item Sample $\mathbf{z}$ from the set $\{\mathbf{z_j}\}_{j=1}^{n}$ weighed by $\{w_i\}_{i=1}^{n}$
    \item Sample $\mathbf{x_m} $ from $p(\mathbf{x_m} |  \mathbf{z})$
\end{enumerate}

Note that conditional on a fixed sample of $\{\mathbf{z_j}\}_{j=1}^{n}$, $p(\mathbf{x_m} | \mathbf{x_o})$ is a mixture of Gaussians with weights $\{w_i\}_{i=1}^{n}$. Using this simple observation, we can see that the point estimate can be approximated as:
\begin{equation}\label{eqn:mean_imp}
    \EE{\mathbf{x_m} | \mathbf{x_o}} \approx \sum_{i=1}^{n} w_i \EE{\mathbf{x_m} | \mathbf{z_i}} 
\end{equation}
where $\EE{\mathbf{x_m} | \mathbf{z}}$ in the case of \myref{Equation}{eqn:vanilla_gen} is simply $\mu_\theta(\mathbf{z_i})_m$. The main value of this approach is to reduce the variance of the estimator.

Similar to how we can approximate the mean, we can also approximate the variance as:
$$ \VV{\mathbf{x_m} | \mathbf{x_o}} \approx \sum_{i=1}^{n} w_i \left(\VV{\mathbf{x_m} | \mathbf{z_i}} - \EE{\mathbf{x_m} | \mathbf{z_i}}\right)^2 - \EE{\mathbf{x_m} | \mathbf{x_o}}^2 $$

\subsection{Practical Implementation}\label{sec:dist_imp_practical}

The algorithm we describe in the previous section was used by \citet{mattei2019miwae} to impute with VAEs. However, we found that there is a simple change we can make to improve this further.

In \citet{mattei2019miwae}, the imputation goal was, given a set of data with missing variables, fill in the gaps. However, our goal is to train a VAE on some universe of data (which may and may not contain missing values) and to impute surfaces on another dataset, i.e., train on a training set and then impute on a test set. This seemingly small difference requires another step to be included into the imputation algorithm in order to get accurate point estimates.

The efficiency of the above algorithm rests upon the accuracy of $q(\mathbf{z} | \mathbf{x_o})$ which we noted to be the posterior. To achieve an accurate estimate of the posterior, we can re-optimize (refit) the encoder $q(\mathbf{z} | \mathbf{x_o})$ \citep{Mattei2018RefitYE} using:
$$ \argmin_{\phi} \mathcal{L}_{\text{VAE}}(\mathbf{x}_o; \sigma_x, \theta, \phi) $$
Note, that we only optimize $\phi$, we keep $\theta$ fixed, or, in other words, we freeze our generative model and re-optimize the posterior estimate. This is valid to do since, given $\theta$, the definition of the posterior is fixed, i.e.,
$$ p(\mathbf{z} | \mathbf{x}) \propto p_\theta(\mathbf{x} |\mathbf{z})\ p(\mathbf{z}) $$
and the VAE loss (or the ELBO) is simply a way to get an accurate estimate of the posterior when $\theta$ is fixed. This simple change leads to a major improvement in performance.

\subsection{Comparison to Prior Work}\label{sec:comparison_vae_imp}

As mentioned in \myref{Section}{sec:related_work}, prior approaches to imputation with VAEs were based on MCMC \citep{rezende2014stochastic,mattei2018dlvm}. This approach was also used by \citet{richert2022vaeswaption} to impute swaption cubes. 
The algorithm can be summarized as:
\begin{enumerate}
    \item Randomly impute $\mathbf{x}$ (or impute with zeros)
    \item Sample $\mathbf{z}$ from $q(\mathbf{z} | \mathbf{x})$
    \item Sample $\hat{\mathbf{x}}$ from $p(\mathbf{x} | \mathbf{z})$
    \item Update the imputed values in $\mathbf{x}$ given the values in $\hat{\mathbf{x}}$
    \item Repeat Steps 2-4 until convergence
\end{enumerate}
One main assumption for convergence of this algorithm is that $q(\mathbf{z} | \mathbf{x})$ is an accurate estimate of the posterior.
However, another problem is that you need to run both the encoder and decoder as many times as you want samples whereas the algorithm in \myref{Section}{sec:dist_imp_algo} requires the encoder to be run once and the decoder as many times are you want samples. Further, for effective MCMC, if we want $N$ samples, it is often encouraged to have a burn-in period (drop the first $B$ samples) and further, keeping only every $K$ samples \citep{brooks2011MCMChandbook}, i.e., requiring $B + N K$ calls to the encoder and decoder.

In other works such as \citet{bergeron2022variational} and \citet{ sokol2022amm}, the $\mathbf{z}$ that optimally reconstructs the observed data is found and used for imputation. This is analogous to refitting the encoder using the reconstruction loss as opposed to the VAE loss (as used in \myref{Section}{sec:dist_imp_algo}. Further, the distributional imputation algorithm optimizes for a distribution over $\mathbf{z}$ instead of a single $\mathbf{z}$; thus, instead of using the imputed values from a single $\mathbf{z}$, the imputed value is a weighted average across multiple $\mathbf{z}$ (\myref{Equation}{eqn:mean_imp}).

\section{Experiments}\label{sec:experiments}

\subsection{Experimental Methodology}

\subsubsection{Data}\label{sec:data_desc}

Similar to \citet{bergeron2022variational}, we evaluate our imputation methodology on FX options. Specifically, we use option surfaces for AUDUSD, USDCAD, EURUSD, GBPUSD, and USDMXN and used 40 points from each surface: five deltas (0.1, 0.25, 0.5, 0.75, and 0.9) and eight tenors (one week, one month, two months, three months, six months, nine months, one year, and three years). We will report our evaluation metrics on the AUDUSD option surface in the body of this paper and leave evaluation metrics on the other surfaces in \myref{Appendix}{sec:mae_all_surfaces} and \myref{Appendix}{sec:mae_tenors_all_surfaces}.

We used data from the beginning of 2012 to the end of 2022. One major difference in our experimental setup is that we partitioned our data into three sets as opposed to two: training set (2012-01-01 to 2020-02-29), validation set (2020-03-01 to 2020-12-31), and test set (2021-01-01 to 2022-12-31). We will report our evaluation metrics on the test set in the body of this paper and leave evaluation metrics on the validation set in \myref{Appendix}{sec:mae_all_surfaces} and \myref{Appendix}{sec:mae_tenors_all_surfaces}.

For evaluation, we randomly mask some percentage of the data and compute the mean absolute error (MAE) between the imputed values and the observed values. Another important difference in our work is that we report two metrics: the MAE on the missing data and the MAE on the observed data, as opposed to the MAE on the full surface. The second metric gives insight into how well the imputed surface faithfully reflects observed market data. Similar to the performance on the missing data, we only report the metrics on AUDUSD in the body of this paper and leave the rest to  \myref{Appendix}{sec:mae_obs_all_surfaces}.

\subsubsection{Baselines}

Prior work \citep{bergeron2022variational} compared imputation performance between VAEs and Heston model \citep{Heston1993ACS}. We introduce two additional baselines to further illustrate the efficacy of VAEs: Heston with jumps \citep{Bates2015HestonJumps} and Heston with shifted volatility (Heston++, \cite{pacati2014hestonpp}).

\subsubsection{Model Details}

For model training, we tuned our hyperparameters on the ELBO in the validation set; similar to the observation in \myref{Section}{sec:dist_imp_practical} that we can improve the posterior estimate by refitting the encoder, we refit our encoder on the validation set to get an accurate estimate of the validation ELBO. We specifically tuned the hidden size (64 and 128), the number of layers (2, 3, and 4), dropout rate (0.1, 0.2, 0.3, and 0.4), and the embedding size (32 and 64). We tune our models with respect to the validation ELBO; we show in \myref{Section}{sec:hparam_tuning} that tuning hyperparameters on the validation set correlates well with performance on the test set. 

For imputation, we compute the mean estimate using the algorithm from \myref{Section}{sec:imputation} using 10K samples from the encoder. Additional details of architecture, training and imputation methodology can be found in \myref{Appendix}{sec:exp_detail_fx}.

\subsection{Point Estimate Results}

\begin{figure}[!bt]
\begin{minipage}{\linewidth}
 \centerline{\includegraphics[width=0.7\linewidth]{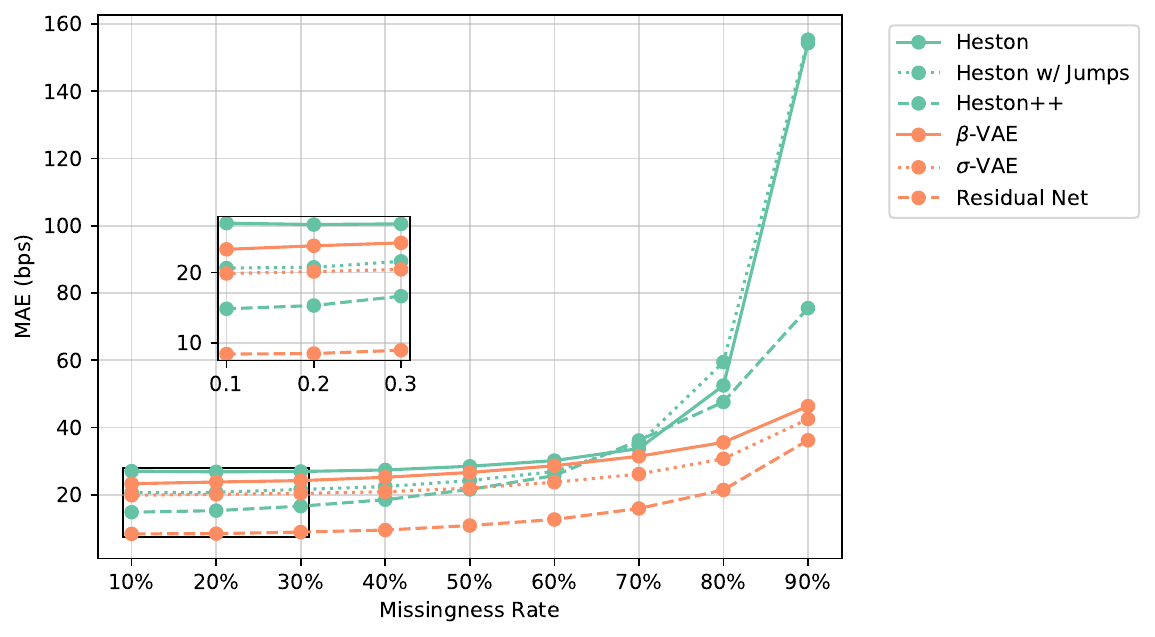}}
    \caption{Mean absolute error (MAE) measured in bps on the test between imputed and masked out data for varying levels of missingness. $\sigma$-VAE uses VAE with traditional MLP blocks and ``Residual Net'' is a $\sigma$-VAE with residual blocks.}
\label{fig:audusd_mae_test_full}
\end{minipage}
\begin{minipage}{\linewidth}
    \centerline{\includegraphics[width=\linewidth]{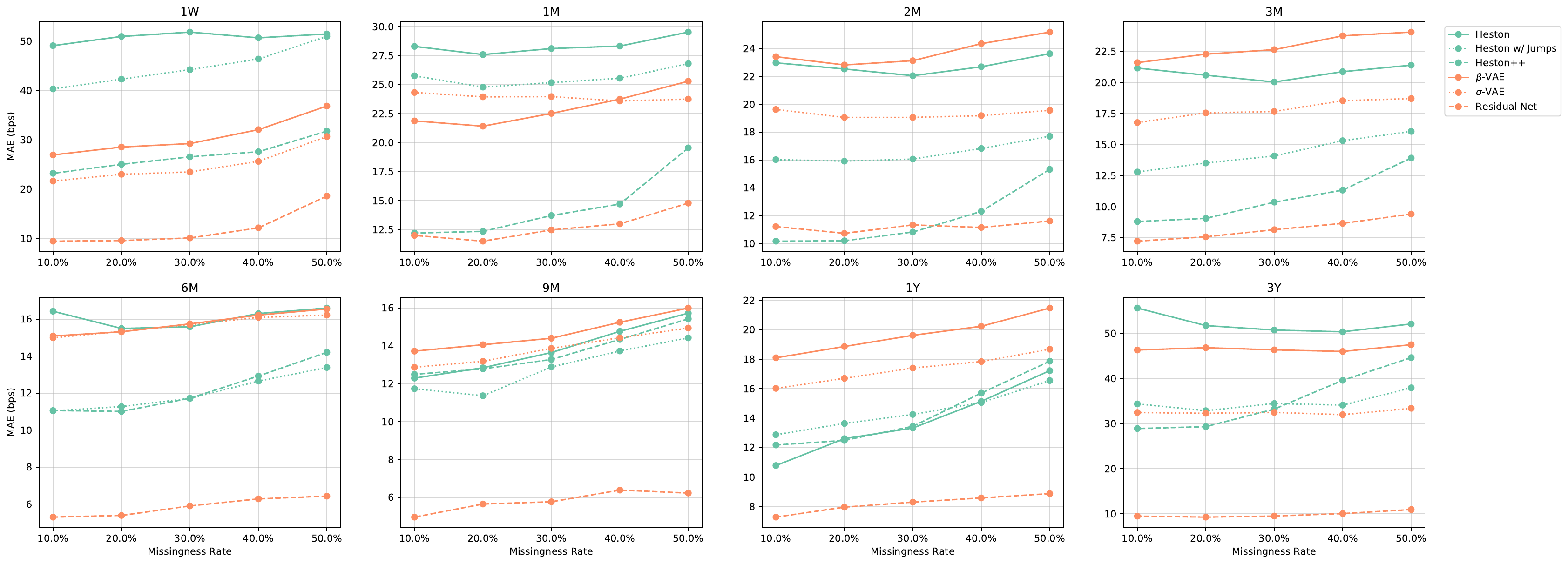}}
    \caption{Mean absolute error (MAE) measured in bps on the test between imputed and masked out data for varying levels of missingness and different tenors. $\sigma$-VAE uses VAE with traditional MLP blocks and ``Residual Net'' is a $\sigma$-VAE with residual blocks.}\label{fig:audusd_mae_test_tenor}
\end{minipage}
\end{figure}

\subsubsection{$\sigma$-VAE and Residual Blocks}\label{sec:perf_resid}

For our first set of results, we compare the three baseline models against $\beta$-VAE ($\beta=1$) with traditional MLP blocks, $\sigma$-VAE with traditional MLP blocks, and $\sigma$-VAE with residual connections. First, in \myref{Figure}{fig:audusd_mae_test_full}, we see that Heston is outperformed by Heston with Jumps and Heston++. Further, Heston++ outperforms across all missingness rates except around 70\%. Next, we see that $\sigma$-VAE outperforms $\beta$-VAE; similar to our observation in \myref{Section}{sec:beta_vae}, these models are equivalent for a specific choice of $\beta$ given $\sigma$ with the key difference that the $\sigma$ in $\sigma$-VAE did not need to be tuned.

Similar to previous findings \citep{bergeron2022variational}, the $\sigma$-VAE outperforms Heston with the largest gap in performance at higher rates of missingness; this observation holds true for Heston with jumps. However, deeper analysis in \myref{Figure}{fig:audusd_mae_test_tenor} where the results are broken down by tenors makes it clear that the $\sigma$-VAE outperforms Heston with jumps due to the smaller tenors (e.g., one week and one month) but is severely outperformed in the two, three, and six month tenors. 

Further, Heston++ outperforms $\sigma$-VAE for missingness rates less than 50\% (\myref{Figure}{fig:audusd_mae_test_full}), with the gap being much larger in certain tenors (\myref{Figure}{fig:audusd_mae_test_tenor}).

Finally, using residual blocks significantly outperforms all other models across missingness rates with the MAE dropping by more than half for 10\% missingness rate (19.9 bps without residual blocks to 8.4 bps with residual blocks, \myref{Figure}{fig:audusd_mae_test_full}). Further, at the tenor level, the residual network outperforms all models except with the Heston++ slightly outperforming in the two months tenor at the 10-30\% missingness rate.

\subsubsection{$\Sigma$-VAE and IWAE}\label{sec:perf_iwae}

\begin{figure}[!bt]
\begin{minipage}{\linewidth}
 \centerline{\includegraphics[width=0.7\linewidth]{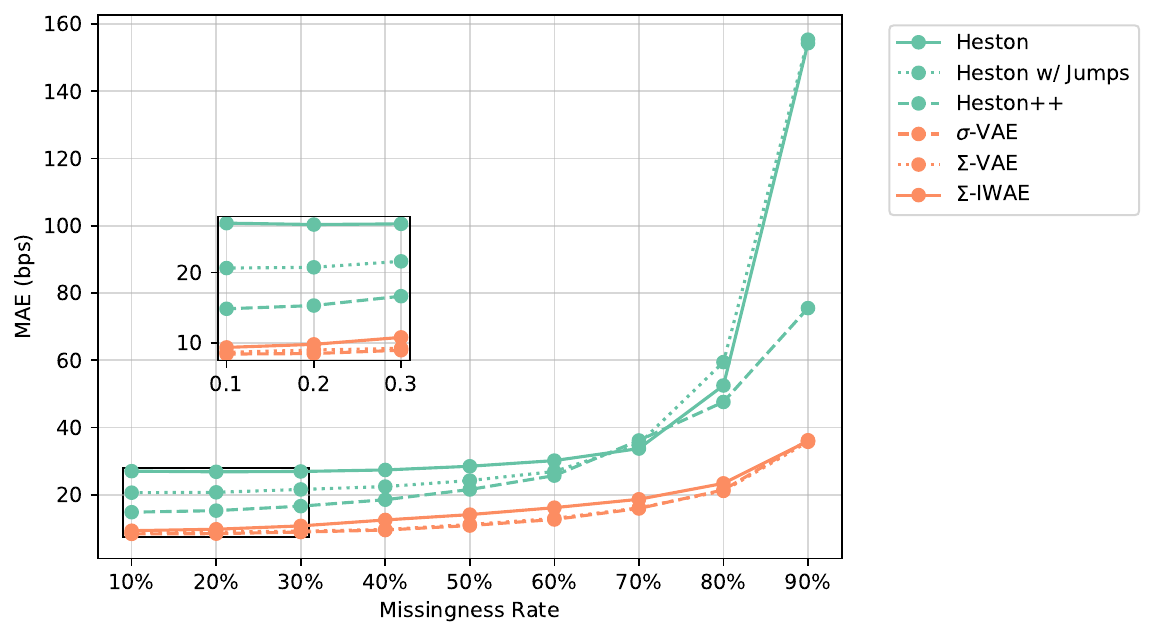}}
    \caption{Mean absolute error (MAE) measured in bps on the test between imputed and masked out data for varying levels of missingness. All VAEs use residual blocks.}
\label{fig:audusd_Sigma_mae_test_full}
\end{minipage}
\end{figure}

For our second set of results, we compare the  best model in the previous section ($\sigma$-VAE with residual blocks) with $\Sigma$-VAE (where the noise $\sigma_x$ is a function of $z$) and $\Sigma$-IWAE. In \myref{Figure}{fig:audusd_Sigma_mae_test_full}, we see that the imputation performance of $\Sigma$-VAE is nearly identical to $\sigma$-VAE; while we might expect the performance to improve, the main intention of using $\Sigma$-VAE is to improve the uncertainty estimates which we analyze in the next section. Interestingly, $\Sigma$-IWAE performs worse in terms of imputation performance; since IWAE loss has less bias than VAE loss, we suspect the degradation in performance might be from overfitting. Future work might explore hyperparameter tuning more thoroughly to improve the IWAE.

Further, in \myref{Appendix}{sec:mae_all_surfaces}, we see that, while the $\Sigma$-VAE is comparable to $\sigma$-VAE on AUDUSD, $\Sigma$-VAE performs similar to $\sigma$-VAE without residuals in low missingness regimes for USDMXN in the validation set, but much better (similar to $\sigma$-VAE with Residuals) in the test set.

\subsubsection{Reconstruction of Observed Data}

\begin{figure}[!bt]
\begin{minipage}{\linewidth}
 \centerline{\includegraphics[width=0.7\linewidth]{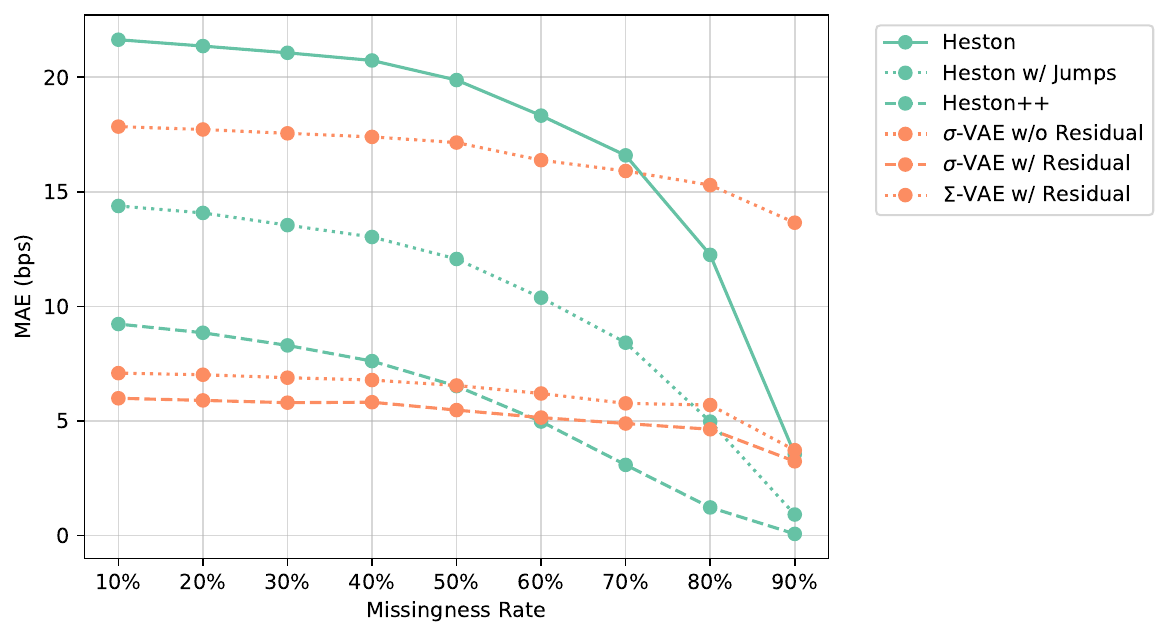}}
    \caption{Mean absolute error (MAE) measured in bps on the test between imputed and observed data for varying levels of missingness.}
\label{fig:audusd_observed_data_full}
\end{minipage}
\end{figure}

Along with the ability to fill in missing data, a requirement of calibrating to market data is to be able to faithfully reproduce the observations from the market. In \myref{Figure}{fig:audusd_observed_data_full}, we compare the MAE in bps on the observed data for our baselines and three VAE variants. In the low missingness regimes, we can see that the VAEs with residual networks is better able to reproduce the observed data. The classical models performs worse because they do not have a sufficient
number of parameters to fit the data.

Interestingly, in the high missingness regimes, the classical models (specifically Heston++) has a significantly lower error rate. This is to be expected since the fitting procedure is aimed at reconstructing the observed data as accurately as possible and the classical models have enough parameters to fit perfectly. The reason the VAE does not have a lower error rate is because of our distributional imputation algorithm. The goal of the algorithm is to not perfectly reconstruct the observed data, but to infer the posterior that balances the prior as well as the reconstruction term. The trade-off between the two terms is a function of the learned $\sigma$, i.e., the noisiness of the data: the more noisy the data (the higher the $\sigma$), the less accurate of a reconstruction. This can be interpreted as the distributional imputation algorithm \textit{regularizing} the imputation to balance the fact that there is not a lot of observed data.

\subsection{Distributional Results}

\begin{table}[!b]
\begin{center}
\begin{tabular}{lcc}
\toprule
     & \multicolumn{2}{c}{Negative ELBO ($\downarrow$)}
\\
 \cmidrule(lr){2-3} 
{Model} & {Train} & {Validation}
\\
\midrule
$\sigma$-VAE (No Residual) & $27.0$ & $39.2$ \\
$\sigma$-IWAE (No Residual) & $12.9$  & $50.4$ \\
$\sigma$-VAE (Residual) & $-6.9$ & {$24.7$} \\
$\sigma$-IWAE (Residual) & $-19.8$  & {$67.6$} \\
$\Sigma$-VAE (Residual) & $-19.1$ & {$\mathbf{9.4}$} \\
$\Sigma$-IWAE (Residual) & $\mathbf{-28.1}$ & {${12.3}$} \\
\bottomrule
\end{tabular}
\captionof{table}{Comparison of Negative ELBO on validation set for different VAEs.}\label{tbl:elbo_comparison}
\end{center}
\end{table}

In \myref{Table}{tbl:elbo_comparison}, we compare the negative ELBO (approximate the negative log likelihood, or, in other words, the distributional performance) for different training methodologies for VAEs \footnote{We do not compare distributional performance against Heston and its variants because it does not give a density function over vol surfaces.} We see that consistently changing to an importance weighted loss hurts performance in the validation set. However, we see the training loss improves; thus, our results are not inconsistent with the fact that IWAEs have less bias and the simulated data example in \myref{Section}{sec:iwae}.

We can see that along with the improvement in imputation that we saw in the previous section for residual networks, we see an improve in the negative ELBO. Further, while there was not an improvement in terms of imputation performance for $\Sigma$-VAE, there was an improvement in terms of negative ELBO.

While negative ELBO is a proper scoring rule (i.e., a metric that reaches its minimum value by only the optimal model) \citep{BrierScore}, we further explore evaluation with a more intuitive metric. Intuitively, we would like the predicted variance to be larger when the model makes larger errors, or, more rigorously, when the model predicts a variance of $\sigma^2$, then
the average squared error should be $\sigma^2$.
In \myref{Figure}{fig:variance_perf}, we evaluate our model against this property. We expect the error-bars to overlap with the $y=x$ line. We have two observations:
\begin{enumerate}
    \item The $\Sigma$-VAE has more variance in its predicted variance than $\vec{\sigma}$-VAE, which has more variance than $\sigma$-VAE. While the $\sigma$-VAE was homoscedastic output noise, there still is a range of predicted variances is because the imputation algorithm from \myref{Section}{sec:dist_imp_algo} (\myref{Equation}{eqn:mean_imp}) integrates over multiple values of $z$.
    \item The predicted variance by $\sigma$-VAE and  $\vec{\sigma}$-VAE is overestimating the true variance; the $\Sigma$-VAE is more accurate with some overestimation near $10^{-7}$.
\end{enumerate}

\myref{Figure}{fig:variance_perf} shows the value of uncertainty estimates: a model trained to output accurate uncertainty estimates can give practitioners better insight into the accuracy of the model. Specifically, the model's predicted variance is a way for the model to convey its prediction of how accurate imputation prediction is.

\begin{figure}[!bt]
\begin{minipage}{\linewidth}
 \centerline{\includegraphics[width=0.7\linewidth]{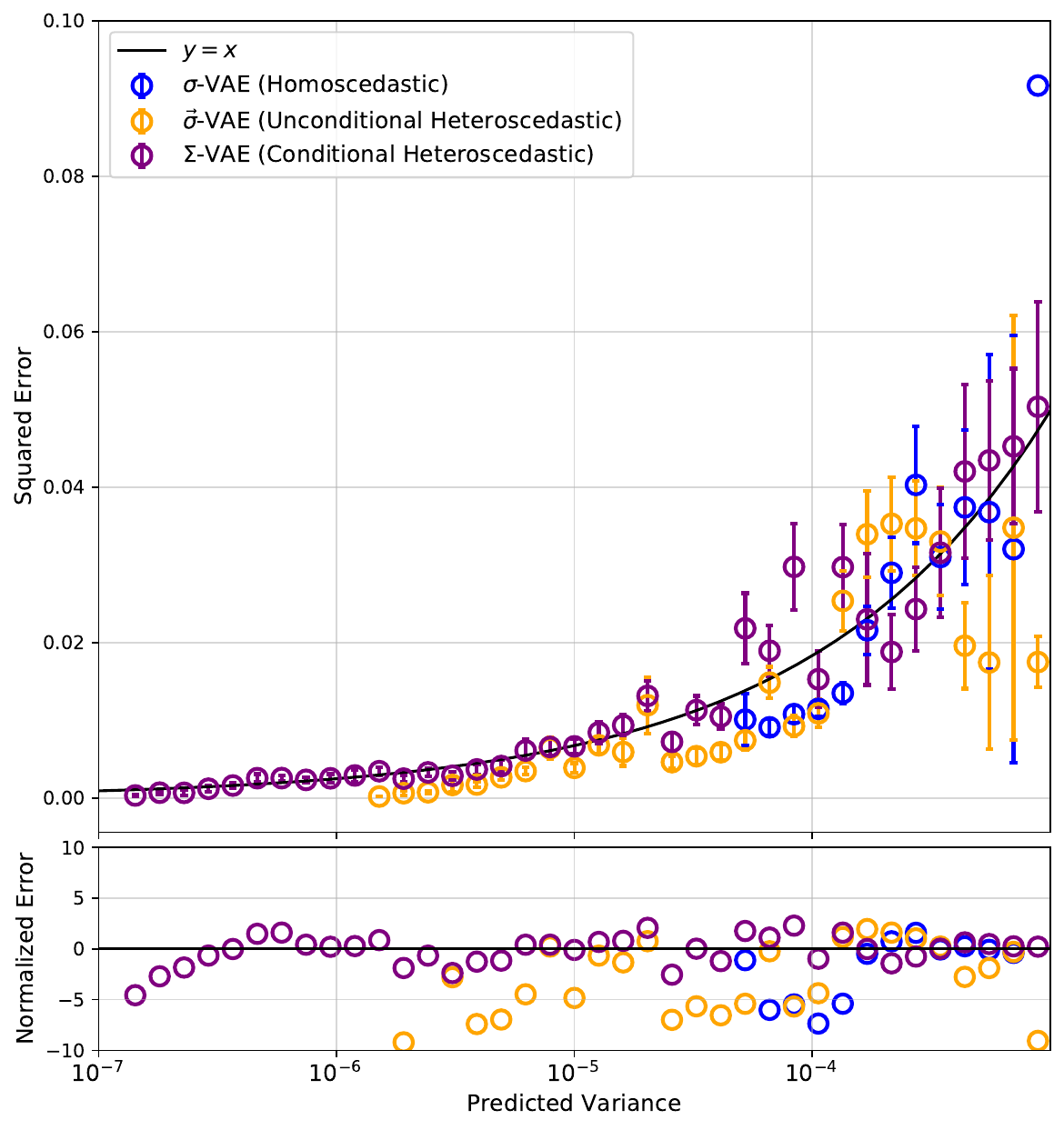}}
    \caption{Performance of variance prediction by VAEs. We compare using a scalar for the output noise (homoscedastic), a vector of the output noise (unconditional heteroscedastic), and noise conditional on $z$ (conditional heteroscedastic). The error-bars in the top plot show the mean and standard error of the squared errors in each bin. The bottom plot shows the distance of the mean from $y=x$ line normalized by the standard error. The $y=x$ line is curved because the $x$-axis is log-scaled.}
\label{fig:variance_perf}
\end{minipage}

\end{figure}

\subsection{Additional Experiments}

\subsubsection{Hyperparameter Tuning}\label{sec:hparam_tuning}
\begin{figure}[!b]
\begin{minipage}{\linewidth}
    
    \begin{subfigure}{0.45\linewidth}
    \centerline{\includegraphics[width=\linewidth]{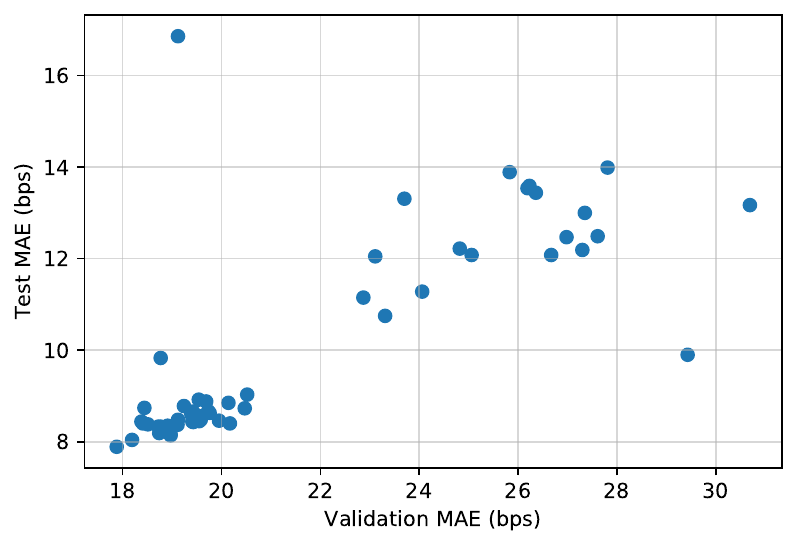}}
    \caption{Comparison of MAE for validation and test}\label{fig:val_mae_v_test_vae}
    \end{subfigure}\hfill
    \begin{subfigure}{0.45\linewidth}
    \centerline{\includegraphics[width=\linewidth]{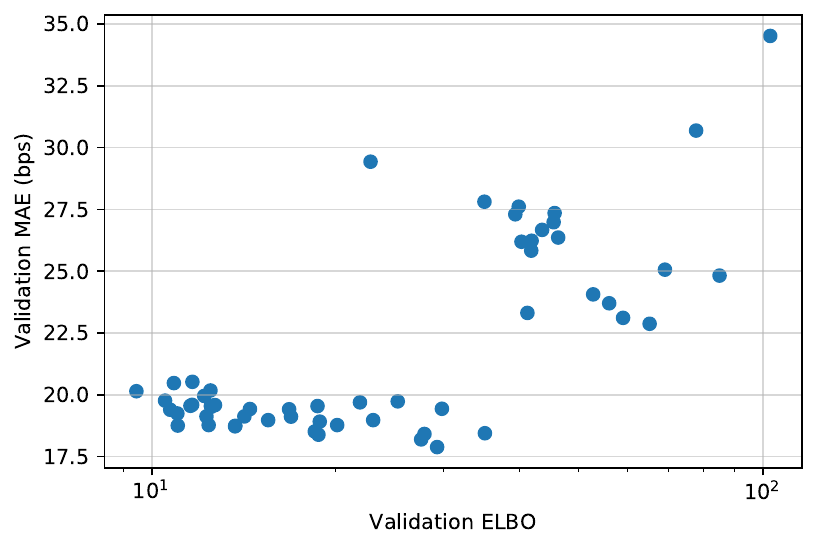}}
    \caption{Comparison of MAE and ELBO for validation }\label{fig:val_elbo_v_val_vae}
    \end{subfigure}
    \caption{Comparison of $\Sigma$-VAE performance across hyperparameters on AUDUSD surfaces.}
\end{minipage}
\end{figure}
As we noted earlier, we hyperparameter tuned all VAEs across 64 different configurations with different hidden sizes (64 and 128), the number of layers (2, 3, and 4), dropout rates (0.1,
0.2, 0.3, and 0.4), and the embedding sizes (32 and 64). A question we asked was, ``How correlated is the performance on the validation set to the performance on the test set?'' In \myref{Figure}{fig:val_mae_v_test_vae}, we compare the performance for AUDUSD with 10\% missingness rate. We see there is a strong correlation (0.77) between the two MAEs, with the optimal model in the validation set being the optimal model for the test set. In other words, \myref{Figure}{fig:val_mae_v_test_vae} implies that we are not overfitting to the validation set. 

Further, we hyperparameter tuned the validation ELBO and accordingly asked, ``Is optimizing for ELBO a good proxy for imputation performance, or, in other words, how correlated is the validation ELBO and validation MAE?'' In \myref{Figure}{fig:val_elbo_v_val_vae}, we see there is a strong correlation (0.77 again) between the two metrics. However, we see that the validation MAE does not seem to improve for ELBO less than 50; this implies that, for validation ELBO, the improvement is coming from the accurate uncertainty estimation. In other words, \myref{Figure}{fig:val_mae_v_test_vae} implies that using the validation ELBO is an effective metric to tune.

\subsubsection{Choosing Embedding Size}

\begin{figure}[!bt]
\begin{minipage}{\linewidth}
    \begin{subfigure}{0.45\linewidth}
    \centerline{\includegraphics[width=\linewidth]{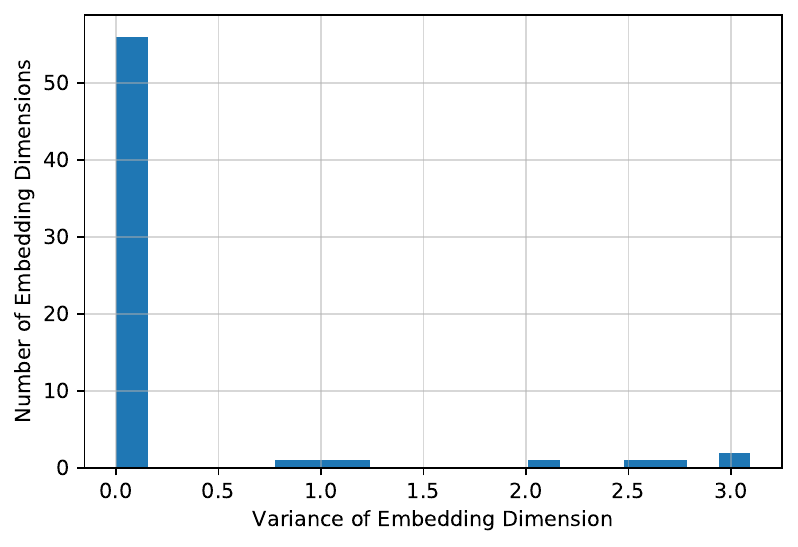}}
    \caption{Histogram of variances}\label{fig:histogram_of_variance}
    \end{subfigure}\hfill
    \begin{subfigure}{0.45\linewidth}
    \centerline{\includegraphics[width=\linewidth]{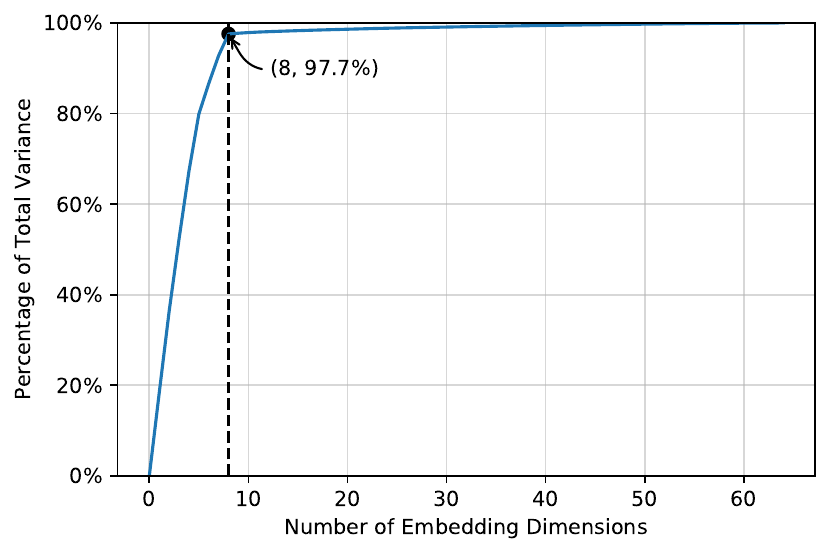}}
    \caption{Skree plot of variances of embeddings}\label{fig:elbow_plot_variance}
    \end{subfigure}
    \caption{Analysis of posterior collapse in VAEs for $\Sigma$-VAE where the embedding dimension is 64.}\label{fig:number_of_embeddings}
\end{minipage}
\end{figure}

A common question when using VAEs is ``how many embedding dimensions to use?'' The embedding size tends to be a common hyperparameter that is tuned (e.g., \cite{bergeron2022variational, sokol2022amm}). We show in \myref{Figure}{fig:number_of_embeddings} that the number of embedding dimensions is automatically tuned by the VAE. Prior works (e.g., \cite{burda2015iwae,Vahdat2020NVAE,Wallach2019DontBlameELBO}) have discussed the problem of ``posterior collapse'' where some latent dimensions are independent of the inputs. In other words, the mean of the posterior for some latent dimensions is zero for all samples (assuming $\mathcal{N}(0, I)$ prior). To evaluate the number of dimensions that have collapsed, we compute $\mu_{\mathbf{z}|\mathbf{x}}$ for each data point and then take the variance of each dimension. In \myref{Figure}{fig:histogram_of_variance}, we plot the histogram of the variances, and we see there is clear separation between some dimensions being very close to zero and others larger than 0.5. In \myref{Figure}{fig:elbow_plot_variance}, we plot the total variance of the embeddings explained by each dimension and see that 8 explains 97.7\% of the variance in $\mu_{\mathbf{z}|\mathbf{x}}$. In other words, the VAE learned without supervision to use only eight embedding dimensions.

\subsubsection{Arbitrage Evaluation}

\begin{table}[!b]
\centering
\begin{tabular}{lccccc}
\toprule
 & AUDUSD & EURUSD & USDCAD & GBPCAD & USDMXN \\
\midrule
$\sigma$-VAE w/o Residual & $35.8 \pm 0.1$ & $24.5 \pm 0.1$ & $34.6 \pm 0.1$ & $21.6 \pm 0.1$ & $38.6 \pm 0.1$ \\
$\sigma$-VAE w/ Residual & $78.9 \pm 0.1$ & $\mathbf{73.7 \pm 0.1}$ & $87.7 \pm 0.1$ & $65.9 \pm 0.1$ & $\mathbf{76.8 \pm 0.1}$ \\
$\Sigma$-VAE w/ Residual & $\mathbf{81.8 \pm 0.1}$ & ${73.2 \pm 0.1}$ & $\mathbf{91.2 \pm 0.1}$ & $\mathbf{68.2 \pm 0.1}$  & ${72.2 \pm 0.1}$ \\
\midrule
Real & $100.0$ & 100.0 & 100.0 & 99.9 & 99.9 \\
\bottomrule
\end{tabular}
\caption{Percentage of surfaces which contain no-arbitrage opportunities, evaluated on 100K samples.}\label{tbl:arb_checks}
\end{table}

One major component in option pricing and volatility surface construction is ensuring no-arbitrage, i.e., the prices given from the model do not allow for non-zero probability of making money from no money. Many works (e.g., \cite{Davis2007RangeOfOptionPrices,Roper2010}) have characterized the conditions required for an option surface to be arbitrage free; the conditions can be summarized as requiring positive implied volatility, monotonicity in time to maturity (calendar arbitrage condition), convex along the strike dimension (butterfly arbitrage condition), and boundary conditions. 

For this work, since our surface has a discrete number of observations, we use:
\begin{theorem}\label{thm:arb_check}
    A set of call prices $\{p_{k, t}\}$ are consistent with no arbitrage (assuming the forward price is equal for all $t$) if and only if
    \begin{itemize}
        \item $K_{k,t} = K_{k',t'}$ and $t < t'$ implies $p_{k, t} < p_{k', t'}$
        \item $\forall k, \forall t$,
        $$ \sup_{t' \geq t} \sup_{\{k': K_{k',t'}  < K_{k,t}  \}} \frac{p_{k, t} - p_{k', t'}}{K_{k,t}  - K_{k',t'} } \leq \min\left(0,  \inf_{t'' \geq t} \inf_{\{k'': K_{k'',t''}  > K_{k,t}  \}} \frac{p_{k'', t''} - p_{k, t} }{K_{k'',t''}  - K_{k,t} } \right) $$
    \end{itemize}
\end{theorem}
where we use $K_{k,t}$ to denote strike at a given strike-index $k$ and tenor $t$. 
\myref{Theorem}{thm:arb_check} gives us a simple algorithm to run on generated volatility surfaces (converted to call prices space) to test for arbitrage. 

In \myref{Table}{tbl:arb_checks}, we compute the percentage of surfaces generated by different VAEs that contain an arbitrage. By construction, our baseline models are guaranteed to be no arbitrage and thus, we do not include them in our table. For each model, we sample 100K times (using the algorithm explained in \myref{Section}{sec:dlvm}) and compute the percentage of which have no-arbitrage. We can see that there's a massive improvement by including residual connections and some benefits by using $\Sigma$-VAE over $\sigma$-VAE, but the result is not consistent across all currency pairs. Note that we do not include any constraints during training to enforce no-arbitrage conditions, the performance we see is from what the model learned from the data. Further, we note that the easiest way to perform well on this task is to simply memorize the training set. However, this form of overfitting would prevent the model from being able to perform accurate imputations on the validation and test set.

\section{Conclusion}

In this work, we introduce a simple set of tricks to improve the imputation performance of VAEs on FX volatility surfaces. With these, we close the gap between the usage of VAEs in imputing volatility surfaces and the usage of VAEs for imputation in the machine learning community. Namely, we show using residual networks leads to stronger performance, $\beta$-VAE is not necessary for training strong VAEs, and VAEs can give accurate uncertainty estimates. Further, we show that using stronger baselines can help discover gaps in performance and accurately measuring the efficacy of an approach requires thorough comparison against classical approaches.

\newpage
\bibliographystyle{chicago}
\bibliography{sample}

\newpage
\appendix

\section{Sampling Noise}\label{sec:sampling_noise}

In this section, using the models trained in \myref{Section}{sec:beta_vae}, we show the importance of sampling from $p(\mathbf{x}|\mathbf{z})$  (using $\sigma_x$) instead of simply using the mean ($\mathbb{E}[\mathbf{x} | \mathbf{z}]$).
In \myref{Figure}{fig:2d_no_noise_generations}, we show the generations when $\sigma_x$ is set to $0$; as expected, the variance observed in the data is not captured by these generations except with very low beta. 

\begin{figure}[!bt]
\begin{minipage}{\linewidth}
 \centerline{\includegraphics[width=0.3\linewidth]{figures/two_dimensional_gauss_data.png}}
    \caption{Mixture of two Gaussians.}
\end{minipage}
\begin{minipage}{\linewidth}
    \begin{subfigure}{0.24\linewidth}
    \centerline{\includegraphics[width=\linewidth]{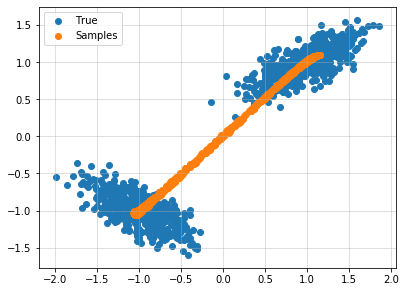}}
    \caption{$\beta=1$}
    \end{subfigure}
    \begin{subfigure}{0.24\linewidth}
    \centerline{\includegraphics[width=\linewidth]{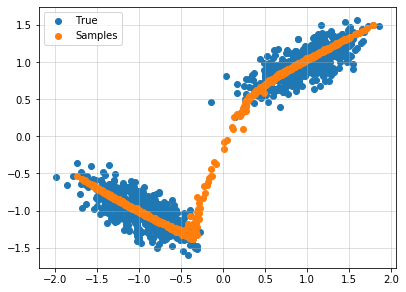}}
    \caption{$\beta=0.1$}
    \end{subfigure}
    \begin{subfigure}{0.24\linewidth}
    \centerline{\includegraphics[width=\linewidth]{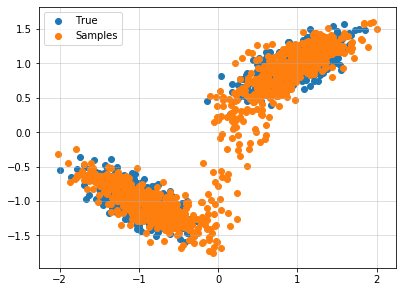}}
    \caption{Learnable $\beta$}
    \end{subfigure}
    \begin{subfigure}{0.24\linewidth}
    \centerline{\includegraphics[width=\linewidth]{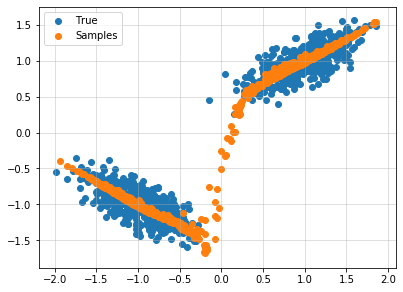}}
    \caption{$\sigma$-VAE}
    \end{subfigure}
    \caption{Comparison of generations by different VAEs with $\sigma_x$ set to zero.}\label{fig:2d_no_noise_generations}
\end{minipage}

\end{figure}

\section{Model Details}

\subsection{VAE Residual Architecture}\label{sec:vae_resid_arch}

While we showed the diagram of our architecture in \myref{Section}{sec:resid}, here, we write the architecture mathematically in order to aid in re-implementing our architecture. Specifically, the VAE architecture in \myref{Figure}{fig:resid_vae} is:
\begin{align*}
    \mathbf{h}_{1, e} &= W_1 \mathbf{x}
\\  \mathbf{h}_{2, e} &= \mathbf{h}_{1, e} + W_3 \sigma(W_2\ \text{LN}(\mathbf{h}_{1, e}))
\\  \mathbf{h}_{3, e} &= \mathbf{h}_{2, e} + W_5 \sigma(W_4\ \text{LN}(\mathbf{h}_{2, e}))
\\ \mu_{\mathbf{z} | \mathbf{x}} = W_{6, 1} \mathbf{h}_{3, e} &\qquad \sigma_{\mathbf{z} | \mathbf{x}} = \text{Softplus}(W_{6, 2} \mathbf{h}_{3, e})
\\ \mathbf{z} &\sim \mathcal{N}(\mu_{\mathbf{z} | \mathbf{x}}, \sigma_{\mathbf{z} | \mathbf{x}})
\\ \mathbf{h}_{1, d} &= W_7 \mathbf{z}
\\ \mathbf{h}_{2, d} &= \mathbf{h}_{1, d} + W_9 \sigma(W_8\ \text{LN}(\mathbf{h}_{1, d}))
\\  \mathbf{h}_{3, d} &= \mathbf{h}_{2, d} + W_{11} \sigma(W_{10}\ \text{LN}(\mathbf{h}_{2, d}))
\\ \mu_{\mathbf{x} | \mathbf{z}} = W_{12, 1} \mathbf{h}_{3, d} &\qquad \sigma_{\mathbf{x} | \mathbf{z}} = \text{Softplus}(W_{12, 2} \mathbf{h}_{3, d})
\end{align*}
where $\mathbf{x} \in \RR^{d}$, $\mathbf{z} \in \RR^{e}$, $W_1 \in \RR^{h \times d}$, $W_2, W_3, W_4, W_5, W_8, W_9, W_{10}, W_{11} \in \RR^{h \times h}$, $W_{6,1}, W_{6, 2} \in \RR^{e \times h}$, $W_7 \in \RR^{h \times e}$, and $W_{12,1}, W_{12, 2} \in \RR^{d \times h}$.
For ease of reading, we dropped all bias terms, $\text{LN}$ refers to Layer Norm \citep{Ba2016LayerNorm}, and $\sigma$ is SiLU \citep{ramachandran2018swish}.

\subsection{Experimental Details for \myref{Section}{sec:beta_vae} and \myref{Section}{sec:iwae}}\label{sec:exp_detail_beta_vae}

For the data, we sampled from a mixture of two Gaussians with equal probability: $$\mathcal{N}\left(\begin{bmatrix}
1.0  \\
1.0
\end{bmatrix}, \begin{bmatrix}
0.1 & 0.05 \\
0.05 & 0.05 
\end{bmatrix}
\right) \qquad \mathcal{N}\left(\begin{bmatrix}
-1.0  \\
-1.0
\end{bmatrix}, \begin{bmatrix}
0.1 & -0.05 \\
-0.05 & 0.05 
\end{bmatrix}
\right) $$
We sampled 1K data points for training.

For our architecture, we used residual networks where, using the notation in \myref{Section}{sec:vae_resid_arch}, $d=2$, $h=64$, and $e=4$. For training, we used the Adam optimizer \citep{Adam2015} with a learning rate of 1e-4 and no weight decay and trained with a batch size of 64 for 100K updates (gradient steps). 
However, since the experiments in \myref{Section}{sec:beta_vae} are for $\sigma-VAE$ and $\beta$-VAE, the layer for $\sigma_{\mathbf{x} | \mathbf{z}}$ is not used.

\subsection{Experimental Details for \myref{Section}{sec:hetero}}\label{sec:exp_detail_hetero_vae}

For the data, we sampled from a mixture of two Gaussians with equal probability: $$\mathcal{N}\left(\begin{bmatrix}
1.0  \\
1.0
\end{bmatrix}, \begin{bmatrix}
0.4 & 0.05 \\
0.05 & 0.05 
\end{bmatrix}
\right) \qquad \mathcal{N}\left(\begin{bmatrix}
-1.0  \\
-1.0
\end{bmatrix}, \begin{bmatrix}
0.4 & -0.05 \\
-0.05 & 0.05 
\end{bmatrix}
\right) $$
We sampled 1K data points for training.

For our architecture, we used residual networks where, using the notation in \myref{Section}{sec:vae_resid_arch}, $d=2$, $h=64$, and $e=4$. For training, we used the Adam optimizer \citep{Adam2015} with a learning rate of $1e{-4}$, no weight decay, and trained with a batch size of 64 for 100K updates (gradient steps).

\subsection{Experimental Details for \myref{Section}{sec:resid}}\label{sec:exp_detail_resid_vae}

For the data, we sampled from a mixture of eight Gaussians with equal probability using the code from the \href{https://github.com/rtqichen/residual-flows/blob/master/resflows/toy_data.py}{residual flow codebase}.
We sampled 1K data points for training.

For our architecture, we used residual networks where, using the notation in \myref{Section}{sec:vae_resid_arch}, $d=2$, $h=64$, and $e=4$. For training, we used the Adam optimizer \citep{Adam2015} with a learning rate of $1e{-4}$, no weight decay, and trained with a batch size of 64 for 100K updates (gradient steps). 

\subsection{Model Details for \myref{Section}{sec:experiments}}\label{sec:exp_detail_fx}

The data we used is described in \myref{Section}{sec:data_desc}.

For each model, we tuned the hidden size (64 and 128), the number of layers (2, 3, and 4), dropout rate (0.1, 0.2, 0.3, and 0.4), and the embedding size (32 and 64). For training, we used the Adam optimizer \citep{Adam2015} weight decay of 1e-5 and trained with a batch size of 64 for 100K updates (gradient steps). For the learning rate, we used a linear warmup from $1e{-7}$ to $2e{-4}$ over 5K steps and then halved the learning rate after 50K steps.

For evaluation, we retrain the encoder on the corresponding dataset with the Adam optimizer \citep{Adam2015} with no weight decay and with a learning rate of $2e{-4}$ where we used a linear warmup for 100 steps. We then trained for 10K steps with a batch size 32 and setting k=50 (even when we trained our model with $k=1$). To remove sources of randomness, we set the dropout rates to zero. Finally, for imputation, using the algorithm in \myref{Section}{sec:imputation}, we use 10K importance weighting samples.

\section{MAE on All Surfaces}\label{sec:mae_all_surfaces}

\begin{table}
    \begin{subtable}{\textwidth}
    \smaller
    \centering
    \begin{tabular}{lrrrrrrrrr}
    \toprule
     & 10\% & 20\% & 30\% & 40\% & 50\% & 60\% & 70\% & 80\% & 90\% \\
    \midrule
    Heston & 27.05 & 26.85 & 26.95 & 27.41 & 28.50 & 30.19 & 33.77 & 52.52 & 154.28 \\
    Heston w/ Jumps & 20.65 & 20.76 & 21.61 & 22.46 & 24.25 & 26.95 & 35.16 & 59.47 & 155.31 \\
    Heston++ & 14.85 & 15.31 & 16.64 & 18.56 & 21.60 & 25.72 & 36.18 & 47.61 & 75.53 \\
    $\sigma$-VAE w/o Residual & 19.88 & 20.15 & 20.46 & 20.92 & 22.01 & 23.73 & 26.18 & 30.72 & 42.52 \\
    $\sigma$-VAE w/ Residual & \textbf{8.40} & \textbf{8.47} & \textbf{8.94} & \textbf{9.54} & \textbf{10.87} & \textbf{12.68} & \textbf{15.95} & 21.47 & 36.24 \\
    $\Sigma$-VAE w/ Residual & 8.65 & 8.97 & 9.21 & 9.71 & 11.18 & 12.99 & 16.13 & \textbf{21.18} & \textbf{35.77} \\
    \bottomrule
    \end{tabular}
    \caption{AUDUSD}
    \end{subtable}
    
    \begin{subtable}{\textwidth}
    \smaller
    \centering
    \begin{tabular}{lrrrrrrrrr}
    \toprule
     & 10\% & 20\% & 30\% & 40\% & 50\% & 60\% & 70\% & 80\% & 90\% \\
    \midrule
    Heston & 20.81 & 20.75 & 20.94 & 21.40 & 22.35 & 23.52 & 28.10 & 47.26 & 150.52 \\
    Heston w/ Jumps & 20.39 & 20.86 & 21.25 & 22.48 & 23.38 & 27.08 & 34.18 & 59.28 & 151.24 \\
    Heston++ & 12.85 & 13.06 & 14.03 & 14.93 & 16.96 & 21.95 & 29.46 & 41.47 & 70.09 \\
    $\sigma$-VAE w/o Residual & 15.36 & 15.47 & 15.91 & 16.41 & 17.04 & 18.54 & 20.69 & 24.67 & 37.31 \\
    $\sigma$-VAE w/ Residual & \textbf{7.50} & \textbf{7.54} & \textbf{7.97} & \textbf{8.75} & \textbf{9.97} & \textbf{12.25} & \textbf{15.57} & \textbf{20.08} & \textbf{34.74} \\
    $\Sigma$-VAE w/ Residual & 8.48 & 8.31 & 8.72 & 9.56 & 11.01 & 12.29 & 15.97 & 20.60 & 37.31 \\
    \bottomrule
    \end{tabular}
    \caption{EURUSD}
    \end{subtable}
    
    \begin{subtable}{\textwidth}
    \smaller
    \centering
    \begin{tabular}{lrrrrrrrrr}
    \toprule
     & 10\% & 20\% & 30\% & 40\% & 50\% & 60\% & 70\% & 80\% & 90\% \\
    \midrule
    Heston & 20.03 & 20.54 & 20.96 & 21.11 & 21.70 & 22.69 & 25.31 & 44.07 & 153.74 \\
    Heston w/ Jumps & 17.20 & 18.40 & 19.28 & 20.62 & 22.30 & 26.34 & 36.69 & 61.34 & 162.48 \\
    Heston++ & 13.56 & 14.26 & 14.87 & 15.86 & 18.09 & 22.17 & 29.87 & 46.28 & 89.90 \\
    $\sigma$-VAE w/o Residual & 16.25 & 16.53 & 16.63 & 16.89 & 17.51 & 18.98 & 20.69 & 24.18 & 33.49 \\
    $\sigma$-VAE w/ Residual & \textbf{6.10} & \textbf{6.57} & 6.94 & 7.62 & 8.68 & 10.65 & 13.54 & 18.36 & \textbf{28.46} \\
    $\Sigma$-VAE w/ Residual & 6.32 & 6.69 & \textbf{6.87} & \textbf{7.24} & \textbf{8.42} & \textbf{9.76} & \textbf{12.26} & \textbf{17.16} & 28.98 \\
    \bottomrule
    \end{tabular}
    \caption{USDCAD}
    \end{subtable}
    
    \begin{subtable}{\textwidth}
    \smaller
    \centering
    \begin{tabular}{lrrrrrrrrr}
    \toprule
     & 10\% & 20\% & 30\% & 40\% & 50\% & 60\% & 70\% & 80\% & 90\% \\
    \midrule
    Heston & 22.71 & 22.44 & 22.86 & 23.20 & 24.05 & 25.41 & 29.24 & 44.88 & 157.29 \\
    Heston w/ Jumps & 21.11 & 20.82 & 21.64 & 22.39 & 24.26 & 26.87 & 34.85 & 58.74 & 153.90 \\
    Heston++ & 13.52 & 14.18 & 15.41 & 16.93 & 19.87 & 23.73 & 32.63 & 46.79 & 79.29 \\
    $\sigma$-VAE w/o Residual & 17.22 & 17.26 & 17.49 & 17.80 & 18.47 & 19.97 & 21.90 & 25.21 & 34.10 \\
    $\sigma$-VAE w/ Residual & \textbf{6.90} & \textbf{7.39} & \textbf{7.91} & \textbf{8.57} & \textbf{9.78} & \textbf{12.06} & \textbf{15.08} & \textbf{20.34} & \textbf{31.09} \\
    $\Sigma$-VAE w/ Residual & 8.54 & 8.66 & 9.12 & 9.62 & 10.96 & 12.59 & 15.78 & 21.42 & 33.16 \\
    \bottomrule
    \end{tabular}
    \caption{GBPUSD}
    \end{subtable}
    
    \begin{subtable}{\textwidth}
    \smaller
    \centering
    \begin{tabular}{lrrrrrrrrr}
    \toprule
     & 10\% & 20\% & 30\% & 40\% & 50\% & 60\% & 70\% & 80\% & 90\% \\
    \midrule
    Heston & 36.05 & 36.68 & 37.07 & 37.47 & 38.62 & 40.63 & 45.07 & 67.27 & 211.61 \\
    Heston w/ Jumps & 28.72 & 29.63 & 30.18 & 30.20 & 32.56 & 36.58 & 46.28 & 77.08 & 197.17 \\
    Heston++ & 30.39 & 32.33 & 32.82 & 33.78 & 37.86 & 44.11 & 59.06 & 96.79 & 201.93 \\
    $\sigma$-VAE w/o Residual & 48.00 & 47.78 & 48.00 & 48.47 & 49.37 & 51.46 & 53.44 & 57.62 & 67.66 \\
    $\sigma$-VAE w/ Residual & \textbf{12.68} & \textbf{12.89} & \textbf{13.51} & \textbf{14.50} & \textbf{16.21} & \textbf{20.08} & \textbf{26.23} & \textbf{37.15} & \textbf{58.81} \\
    $\Sigma$-VAE w/ Residual & 16.33 & 16.64 & 17.16 & 18.24 & 20.15 & 23.77 & 29.34 & 39.54 & 61.65 \\
    \bottomrule
    \end{tabular}
    \caption{USDMXN}
    \end{subtable}
    
    \caption{Mean absolute error (MAE) measured in bps on the test set for varying levels of missingness for different surfaces.}
    \end{table}
\begin{table}
\begin{subtable}{\textwidth}
\smaller
\centering
\begin{tabular}{lrrrrrrrrr}
\toprule
 & 10\% & 20\% & 30\% & 40\% & 50\% & 60\% & 70\% & 80\% & 90\% \\
\midrule
Heston & 37.69 & 39.70 & 40.77 & 41.84 & 43.78 & 46.50 & 49.17 & 71.64 & 184.60 \\
Heston w/ Jumps & 32.28 & 34.17 & 34.91 & 38.29 & 40.45 & 45.01 & 54.40 & 77.75 & 162.26 \\
Heston++ & 24.96 & 27.21 & 28.05 & 29.53 & 32.92 & 40.36 & 54.64 & 74.41 & 103.35 \\
$\sigma$-VAE w/o Residual & 29.55 & 32.72 & 32.75 & 33.28 & 34.15 & 36.98 & 41.88 & 51.76 & 73.40 \\
$\sigma$-VAE w/ Residual & \textbf{15.61} & \textbf{16.94} & \textbf{16.80} & \textbf{17.57} & \textbf{18.96} & \textbf{21.85} & \textbf{27.65} & \textbf{37.82} & 61.23 \\
$\Sigma$-VAE w/ Residual & 20.25 & 21.52 & 22.24 & 23.21 & 24.14 & 26.29 & 30.87 & 39.00 & \textbf{60.56} \\
\bottomrule
\end{tabular}
\caption{AUDUSD}
\end{subtable}

\begin{subtable}{\textwidth}
\smaller
\centering
\begin{tabular}{lrrrrrrrrr}
\toprule
 & 10\% & 20\% & 30\% & 40\% & 50\% & 60\% & 70\% & 80\% & 90\% \\
\midrule
Heston & 26.46 & 27.35 & 27.59 & 28.45 & 28.67 & 30.84 & 33.73 & 53.53 & 161.72 \\
Heston w/ Jumps & 22.86 & 24.74 & 26.13 & 26.32 & 27.61 & 30.47 & 38.14 & 62.53 & 147.85 \\
Heston++ & 15.48 & 16.48 & 17.13 & 18.78 & 20.46 & 25.13 & 31.73 & 48.63 & 83.33 \\
$\sigma$-VAE w/o Residual & 28.71 & 29.05 & 29.56 & 29.77 & 30.52 & 32.56 & 35.35 & 42.51 & 57.60 \\
$\sigma$-VAE w/ Residual & \textbf{10.97} & \textbf{11.49} & \textbf{11.76} & \textbf{12.77} & \textbf{13.87} & \textbf{17.34} & 23.03 & 33.33 & 51.04 \\
$\Sigma$-VAE w/ Residual & 11.64 & 12.80 & 13.58 & 14.55 & 15.32 & 18.49 & \textbf{22.28} & \textbf{31.77} & \textbf{50.82} \\
\bottomrule
\end{tabular}
\caption{EURUSD}
\end{subtable}

\begin{subtable}{\textwidth}
\smaller
\centering
\begin{tabular}{lrrrrrrrrr}
\toprule
 & 10\% & 20\% & 30\% & 40\% & 50\% & 60\% & 70\% & 80\% & 90\% \\
\midrule
Heston & 26.67 & 28.03 & 27.75 & 28.44 & 28.98 & 31.18 & 34.43 & 55.45 & 177.47 \\
Heston w/ Jumps & 24.68 & 25.90 & 25.81 & 26.83 & 30.38 & 34.23 & 46.11 & 69.07 & 149.83 \\
Heston++ & 16.54 & 17.79 & 18.24 & 19.99 & 22.85 & 28.54 & 34.67 & 54.23 & 99.33 \\
$\sigma$-VAE w/o Residual & 25.51 & 26.84 & 27.43 & 27.59 & 28.26 & 30.16 & 32.43 & 39.64 & 57.21 \\
$\sigma$-VAE w/ Residual & \textbf{10.73} & \textbf{11.43} & \textbf{11.91} & \textbf{13.29} & 14.61 & 18.56 & 23.46 & 33.29 & \textbf{51.70} \\
$\Sigma$-VAE w/ Residual & 10.87 & 12.01 & 12.73 & 13.47 & \textbf{14.36} & \textbf{17.33} & \textbf{20.82} & \textbf{30.23} & 51.80 \\
\bottomrule
\end{tabular}
\caption{USDCAD}
\end{subtable}

\begin{subtable}{\textwidth}
\smaller
\centering
\begin{tabular}{lrrrrrrrrr}
\toprule
 & 10\% & 20\% & 30\% & 40\% & 50\% & 60\% & 70\% & 80\% & 90\% \\
\midrule
Heston & 36.41 & 37.47 & 38.07 & 38.88 & 40.07 & 42.07 & 46.53 & 72.97 & 173.81 \\
Heston w/ Jumps & 30.77 & 32.03 & 33.07 & 34.06 & 37.37 & 41.29 & 50.93 & 76.10 & 162.46 \\
Heston++ & 19.65 & 21.17 & 22.45 & 23.59 & 28.17 & 34.76 & 46.42 & 69.64 & 94.34 \\
$\sigma$-VAE w/o Residual & 29.68 & 30.55 & 31.45 & 31.65 & 32.19 & 34.00 & 37.29 & 43.03 & 57.29 \\
$\sigma$-VAE w/ Residual & \textbf{13.24} & \textbf{13.65} & \textbf{13.97} & \textbf{14.92} & \textbf{15.93} & \textbf{19.56} & \textbf{25.29} & \textbf{34.23} & \textbf{51.81} \\
$\Sigma$-VAE w/ Residual & 18.02 & 19.16 & 19.85 & 20.80 & 22.00 & 25.08 & 29.88 & 38.34 & 55.36 \\
\bottomrule
\end{tabular}
\caption{GBPUSD}
\end{subtable}

\begin{subtable}{\textwidth}
\smaller
\centering
\begin{tabular}{lrrrrrrrrr}
\toprule
 & 10\% & 20\% & 30\% & 40\% & 50\% & 60\% & 70\% & 80\% & 90\% \\
\midrule
Heston & 67.06 & 72.22 & 71.64 & 73.20 & 76.05 & 81.40 & 90.28 & 127.06 & 306.38 \\
Heston w/ Jumps & 58.01 & 62.69 & 61.45 & 63.48 & 66.64 & 74.71 & 95.46 & 147.10 & 309.41 \\
Heston++ & 52.44 & 56.96 & 58.04 & 61.37 & 67.74 & 80.33 & 107.27 & 158.56 & 291.70 \\
$\sigma$-VAE w/o Residual & 45.81 & 47.94 & 48.08 & 49.03 & 51.55 & 56.39 & 63.80 & 80.97 & 124.04 \\
$\sigma$-VAE w/ Residual & \textbf{20.45} & \textbf{21.82} & \textbf{22.40} & \textbf{24.24} & \textbf{26.65} & \textbf{31.01} & \textbf{39.64} & \textbf{57.77} & \textbf{106.03} \\
$\Sigma$-VAE w/ Residual & 48.91 & 57.30 & 59.19 & 61.74 & 63.34 & 68.83 & 76.39 & 94.68 & 138.18 \\
\bottomrule
\end{tabular}
\caption{USDMXN}
\end{subtable}

\caption{Mean absolute error (MAE) measured in bps on the validation set for varying levels of missingness for different surfaces.}
\end{table}

\section{MAE on Observed Data in All Surfaces}\label{sec:mae_obs_all_surfaces}
\begin{table}
    \begin{subtable}{\textwidth}
    \smaller
    \centering
    \begin{tabular}{lrrrrrrrrr}
    \toprule
     & 10\% & 20\% & 30\% & 40\% & 50\% & 60\% & 70\% & 80\% & 90\% \\
    \midrule
    Heston & 21.64 & 21.36 & 21.07 & 20.73 & 19.88 & 18.33 & 16.59 & 12.25 & 3.55 \\
    Heston w/ Jumps & 14.38 & 14.08 & 13.55 & 13.03 & 12.07 & 10.38 & 8.41 & 4.97 & 0.91 \\
    Heston++ & 9.23 & 8.85 & 8.30 & 7.61 & 6.52 & \textbf{4.97} & \textbf{3.08} & \textbf{1.22} & \textbf{0.07} \\
    $\sigma$-VAE w/o Residual & 17.85 & 17.72 & 17.55 & 17.40 & 17.15 & 16.38 & 15.91 & 15.29 & 13.65 \\
    $\sigma$-VAE w/ Residual & \textbf{5.99} & \textbf{5.89} & \textbf{5.80} & \textbf{5.82} & \textbf{5.47} & 5.14 & 4.89 & 4.63 & 3.23 \\
    $\Sigma$-VAE w/ Residual & 7.09 & 7.01 & 6.89 & 6.78 & 6.55 & 6.20 & 5.76 & 5.69 & 3.72 \\
    \bottomrule
    \end{tabular}
    \caption{AUDUSD}
    \end{subtable}
    
    \begin{subtable}{\textwidth}
    \smaller
    \centering
    \begin{tabular}{lrrrrrrrrr}
    \toprule
     & 10\% & 20\% & 30\% & 40\% & 50\% & 60\% & 70\% & 80\% & 90\% \\
    \midrule
    Heston & 17.23 & 17.04 & 16.69 & 16.38 & 15.84 & 14.93 & 13.38 & 10.04 & 3.76 \\
    Heston w/ Jumps & 15.27 & 14.96 & 14.53 & 14.01 & 13.18 & 11.93 & 10.12 & 7.22 & 2.13 \\
    Heston++ & 7.97 & 7.68 & 7.51 & 6.76 & 5.96 & \textbf{4.72} & \textbf{3.11} & \textbf{1.37} & \textbf{0.08} \\
    $\sigma$-VAE w/o Residual & 13.72 & 13.71 & 13.66 & 13.57 & 13.54 & 13.41 & 13.55 & 13.74 & 13.90 \\
    $\sigma$-VAE w/ Residual & \textbf{5.44} & \textbf{5.31} & \textbf{5.26} & \textbf{5.15} & \textbf{5.35} & 4.93 & 4.91 & 4.50 & 3.34 \\
    $\Sigma$-VAE w/ Residual & 6.84 & 6.69 & 6.67 & 6.57 & 6.63 & 6.22 & 6.01 & 5.53 & 4.05 \\
    \bottomrule
    \end{tabular}
    \caption{EURUSD}
    \end{subtable}
    
    \begin{subtable}{\textwidth}
    \smaller
    \centering
    \begin{tabular}{lrrrrrrrrr}
    \toprule
     & 10\% & 20\% & 30\% & 40\% & 50\% & 60\% & 70\% & 80\% & 90\% \\
    \midrule
    Heston & 16.71 & 16.34 & 15.96 & 15.59 & 14.97 & 13.83 & 12.20 & 9.36 & 3.45 \\
    Heston w/ Jumps & 12.44 & 12.08 & 11.72 & 11.31 & 10.66 & 9.82 & 8.77 & 6.87 & 2.60 \\
    Heston++ & 8.35 & 7.85 & 7.36 & 6.77 & 5.85 & 4.43 & \textbf{2.74} & \textbf{1.10} & \textbf{0.09} \\
    $\sigma$-VAE w/o Residual & 14.07 & 13.97 & 13.85 & 13.78 & 13.67 & 13.38 & 13.23 & 13.34 & 13.08 \\
    $\sigma$-VAE w/ Residual & \textbf{4.40} & \textbf{4.33} & \textbf{4.35} & \textbf{4.41} & \textbf{4.26} & \textbf{4.22} & 4.15 & 4.02 & 3.06 \\
    $\Sigma$-VAE w/ Residual & 5.16 & 5.06 & 4.94 & 4.87 & 4.68 & 4.46 & 4.10 & 3.73 & 2.67 \\
    \bottomrule
    \end{tabular}
    \caption{USDCAD}
    \end{subtable}
    
    \begin{subtable}{\textwidth}
    \smaller
    \centering
    \begin{tabular}{lrrrrrrrrr}
    \toprule
     & 10\% & 20\% & 30\% & 40\% & 50\% & 60\% & 70\% & 80\% & 90\% \\
    \midrule
    Heston & 18.64 & 18.47 & 18.17 & 17.65 & 17.05 & 15.99 & 14.32 & 10.59 & 3.09 \\
    Heston w/ Jumps & 15.17 & 14.88 & 14.36 & 13.54 & 12.48 & 11.31 & 9.19 & 6.05 & 1.19 \\
    Heston++ & 8.81 & 8.48 & 7.95 & 7.39 & 6.54 & 5.10 & \textbf{3.23} & \textbf{1.43} & \textbf{0.05} \\
    $\sigma$-VAE w/o Residual & 15.18 & 15.15 & 15.07 & 14.86 & 14.73 & 14.05 & 13.45 & 12.51 & 10.87 \\
    $\sigma$-VAE w/ Residual & \textbf{5.05} & \textbf{5.04} & \textbf{4.96} & \textbf{5.01} & \textbf{4.87} & \textbf{4.76} & 4.53 & 4.31 & 2.98 \\
    $\Sigma$-VAE w/ Residual & 7.17 & 7.13 & 7.12 & 7.04 & 6.84 & 6.58 & 6.47 & 6.25 & 5.28 \\
    \bottomrule
    \end{tabular}
    \caption{GBPUSD}
    \end{subtable}
    
    \begin{subtable}{\textwidth}
    \smaller
    \centering
    \begin{tabular}{lrrrrrrrrr}
    \toprule
     & 10\% & 20\% & 30\% & 40\% & 50\% & 60\% & 70\% & 80\% & 90\% \\
    \midrule
    Heston & 28.66 & 28.24 & 27.60 & 26.96 & 25.84 & 24.10 & 21.05 & 15.40 & 6.03 \\
    Heston w/ Jumps & 19.93 & 19.26 & 18.46 & 17.57 & 16.29 & 14.12 & 10.79 & 6.30 & 1.02 \\
    Heston++ & 18.73 & 17.56 & 16.42 & 15.05 & 12.92 & 9.91 & \textbf{6.37} & \textbf{2.62} & \textbf{0.18} \\
    $\sigma$-VAE w/o Residual & 40.87 & 40.39 & 39.76 & 38.87 & 37.95 & 36.19 & 33.51 & 29.98 & 22.96 \\
    $\sigma$-VAE w/ Residual & \textbf{8.34} & \textbf{8.20} & \textbf{8.07} & \textbf{7.90} & \textbf{7.74} & \textbf{7.47} & 7.16 & 6.98 & 5.26 \\
    $\Sigma$-VAE w/ Residual & 12.63 & 12.62 & 12.71 & 12.70 & 12.60 & 12.45 & 12.31 & 11.78 & 8.59 \\
    \bottomrule
    \end{tabular}
    \caption{USDMXN}
    \end{subtable}
    
    \caption{Mean absolute error (MAE) measured in bps on the test set for observed data for varying levels of missingness for different surfaces.}
    \end{table}
    
\begin{table}
    \begin{subtable}{\textwidth}
    \smaller
    \centering
    \begin{tabular}{lrrrrrrrrr}
    \toprule
     & 10\% & 20\% & 30\% & 40\% & 50\% & 60\% & 70\% & 80\% & 90\% \\
    \midrule
    Heston & 32.67 & 32.17 & 31.39 & 30.87 & 29.77 & 27.49 & 24.12 & 17.40 & 6.15 \\
    Heston w/ Jumps & 24.86 & 24.20 & 23.38 & 22.84 & 21.69 & 19.35 & 15.38 & 9.84 & 2.51 \\
    Heston++ & 16.71 & 15.91 & 15.09 & 14.36 & 12.72 & 10.78 & \textbf{6.82} & \textbf{3.14} & \textbf{0.38} \\
    $\sigma$-VAE w/o Residual & 28.44 & 27.97 & 27.83 & 27.58 & 27.11 & 26.63 & 26.10 & 25.43 & 23.06 \\
    $\sigma$-VAE w/ Residual & \textbf{10.88} & \textbf{10.58} & \textbf{10.29} & \textbf{10.06} & \textbf{9.73} & \textbf{9.21} & 8.49 & 7.30 & 4.89 \\
    $\Sigma$-VAE w/ Residual & 17.64 & 17.20 & 16.96 & 16.70 & 16.83 & 16.61 & 14.99 & 12.12 & 10.13 \\
    \bottomrule
    \end{tabular}
    \caption{AUDUSD}
    \end{subtable}
    
    \begin{subtable}{\textwidth}
    \smaller
    \centering
    \begin{tabular}{lrrrrrrrrr}
    \toprule
     & 10\% & 20\% & 30\% & 40\% & 50\% & 60\% & 70\% & 80\% & 90\% \\
    \midrule
    Heston & 22.24 & 21.75 & 21.16 & 20.44 & 19.75 & 18.44 & 16.59 & 12.39 & 4.14 \\
    Heston w/ Jumps & 16.96 & 16.54 & 16.15 & 15.36 & 14.45 & 12.71 & 10.92 & 6.71 & 2.29 \\
    Heston++ & 9.77 & 9.22 & 8.78 & 8.04 & \textbf{7.02} & \textbf{5.84} & \textbf{3.83} & \textbf{1.66} & \textbf{0.14} \\
    $\sigma$-VAE w/o Residual & 24.99 & 24.81 & 24.55 & 24.48 & 24.24 & 24.28 & 23.49 & 23.07 & 20.91 \\
    $\sigma$-VAE w/ Residual & \textbf{7.47} & \textbf{7.47} & \textbf{7.36} & \textbf{7.22} & 7.33 & 7.32 & 7.35 & 6.97 & 4.52 \\
    $\Sigma$-VAE w/ Residual & 10.63 & 10.53 & 10.43 & 10.28 & 10.21 & 10.18 & 9.34 & 8.17 & 5.08 \\
    \bottomrule
    \end{tabular}
    \caption{EURUSD}
    \end{subtable}
    
    \begin{subtable}{\textwidth}
    \smaller
    \centering
    \begin{tabular}{lrrrrrrrrr}
    \toprule
     & 10\% & 20\% & 30\% & 40\% & 50\% & 60\% & 70\% & 80\% & 90\% \\
    \midrule
    Heston & 22.24 & 21.70 & 21.34 & 20.84 & 20.11 & 18.80 & 17.21 & 12.93 & 4.23 \\
    Heston w/ Jumps & 18.60 & 17.92 & 17.66 & 17.18 & 16.39 & 14.43 & 13.30 & 9.36 & 2.64 \\
    Heston++ & 11.05 & 10.64 & 10.19 & 9.37 & 8.25 & \textbf{6.93} & \textbf{4.38} & \textbf{2.11} & \textbf{0.26} \\
    $\sigma$-VAE w/o Residual & 22.81 & 22.61 & 22.30 & 22.24 & 22.07 & 22.17 & 22.39 & 22.27 & 21.89 \\
    $\sigma$-VAE w/ Residual & \textbf{7.37} & \textbf{7.27} & \textbf{7.25} & \textbf{7.42} & \textbf{7.26} & 7.32 & 7.35 & 6.72 & 4.77 \\
    $\Sigma$-VAE w/ Residual & 9.32 & 9.26 & 9.23 & 9.06 & 8.92 & 8.81 & 8.62 & 7.76 & 4.99 \\
    \bottomrule
    \end{tabular}
    \caption{USDCAD}
    \end{subtable}
    
    \begin{subtable}{\textwidth}
    \smaller
    \centering
    \begin{tabular}{lrrrrrrrrr}
    \toprule
     & 10\% & 20\% & 30\% & 40\% & 50\% & 60\% & 70\% & 80\% & 90\% \\
    \midrule
    Heston & 31.06 & 30.50 & 29.84 & 29.18 & 28.05 & 25.86 & 23.22 & 17.55 & 6.14 \\
    Heston w/ Jumps & 23.26 & 22.61 & 21.82 & 20.89 & 19.72 & 17.57 & 14.17 & 10.01 & 2.83 \\
    Heston++ & 14.03 & 13.70 & 13.23 & 12.60 & 11.33 & 9.54 & \textbf{6.60} & \textbf{3.41} & \textbf{0.28} \\
    $\sigma$-VAE w/o Residual & 27.01 & 26.79 & 26.22 & 26.11 & 25.81 & 24.95 & 23.39 & 21.87 & 17.73 \\
    $\sigma$-VAE w/ Residual & \textbf{8.82} & \textbf{8.66} & \textbf{8.47} & \textbf{8.26} & \textbf{7.92} & \textbf{7.55} & 7.37 & 6.56 & 4.76 \\
    $\Sigma$-VAE w/ Residual & 16.10 & 16.14 & 15.98 & 15.84 & 15.94 & 15.76 & 14.41 & 12.95 & 11.52 \\
    \bottomrule
    \end{tabular}
    \caption{GBPUSD}
    \end{subtable}
    
    \begin{subtable}{\textwidth}
    \smaller
    \centering
    \begin{tabular}{lrrrrrrrrr}
    \toprule
     & 10\% & 20\% & 30\% & 40\% & 50\% & 60\% & 70\% & 80\% & 90\% \\
    \midrule
    Heston & 59.08 & 57.82 & 56.91 & 55.71 & 53.54 & 49.82 & 44.38 & 35.12 & 11.90 \\
    Heston w/ Jumps & 45.86 & 44.10 & 44.15 & 42.38 & 40.42 & 36.69 & 31.75 & 24.08 & 6.97 \\
    Heston++ & 38.19 & 36.53 & 34.91 & 32.61 & 29.74 & 25.08 & 17.70 & 10.06 & \textbf{1.54} \\
    $\sigma$-VAE w/o Residual & 41.27 & 40.81 & 40.47 & 40.03 & 39.12 & 38.47 & 37.56 & 36.58 & 33.25 \\
    $\sigma$-VAE w/ Residual & \textbf{14.43} & \textbf{13.98} & \textbf{13.46} & \textbf{13.00} & \textbf{11.98} & \textbf{11.32} & \textbf{10.14} & \textbf{9.61} & 6.65 \\
    $\Sigma$-VAE w/ Residual & 47.76 & 48.36 & 48.85 & 49.90 & 52.35 & 54.77 & 56.70 & 57.82 & 62.55 \\
    \bottomrule
    \end{tabular}
    \caption{USDMXN}
    \end{subtable}
    
    \caption{Mean absolute error (MAE) measured in bps on the validation set for observed data for varying levels of missingness for different surfaces.}
    \end{table}

\section{MAE for Different Tenors}\label{sec:mae_tenors_all_surfaces}

\begin{table}
    \begin{subtable}{\textwidth}
    \smaller
    \centering
    \begin{tabular}{lrrrrrrrr}
    \toprule
    Tenor & 1W & 1M & 2M & 3M & 6M & 9M & 1Y & 3Y \\
    \midrule
    Heston & 49.09 & 28.29 & 22.98 & 21.18 & 16.43 & 12.30 & 10.78 & 55.62 \\
    Heston w/ Jumps & 40.31 & 25.76 & 16.03 & 12.80 & 11.04 & 11.74 & 12.88 & 34.36 \\
    Heston++ & 23.20 & 12.20 & \textbf{10.17} & 8.81 & 11.06 & 12.50 & 12.18 & 28.90 \\
    $\sigma$-VAE w/o Residual & 21.63 & 24.32 & 19.63 & 16.79 & 14.99 & 12.87 & 16.02 & 32.46 \\
    $\sigma$-VAE w/ Residual & \textbf{9.44} & \textbf{12.00} & 11.22 & \textbf{7.23} & 5.29 & \textbf{4.96} & 7.28 & 9.48 \\
    $\Sigma$-VAE w/ Residual & 9.94 & 12.02 & 13.26 & 7.94 & \textbf{5.28} & 5.29 & \textbf{6.88} & \textbf{8.38} \\
    \bottomrule
    \end{tabular}
    \caption{AUDUSD}
    \end{subtable}
    
    \begin{subtable}{\textwidth}
    \smaller
    \centering
    \begin{tabular}{lrrrrrrrr}
    \toprule
    Tenor & 1W & 1M & 2M & 3M & 6M & 9M & 1Y & 3Y \\
    \midrule
    Heston & 38.62 & 26.30 & 19.04 & 14.18 & 10.45 & 11.74 & 15.97 & 29.87 \\
    Heston w/ Jumps & 40.54 & 24.30 & 14.07 & 10.99 & 10.51 & 14.45 & 15.49 & 32.56 \\
    Heston++ & 18.61 & 11.07 & 9.17 & 8.22 & 9.10 & 9.65 & 11.71 & 25.40 \\
    $\sigma$-VAE w/o Residual & 17.23 & 15.56 & 15.12 & 12.96 & 12.20 & 13.40 & 17.29 & 19.15 \\
    $\sigma$-VAE w/ Residual & \textbf{9.23} & \textbf{7.75} & \textbf{7.29} & 6.47 & \textbf{6.07} & 6.67 & \textbf{7.37} & \textbf{9.16} \\
    $\Sigma$-VAE w/ Residual & 10.54 & 7.95 & 10.82 & \textbf{5.89} & 6.15 & \textbf{6.41} & 8.08 & 11.98 \\
    \bottomrule
    \end{tabular}
    \caption{EURUSD}
    \end{subtable}
    
    \begin{subtable}{\textwidth}
    \smaller
    \centering
    \begin{tabular}{lrrrrrrrr}
    \toprule
    Tenor & 1W & 1M & 2M & 3M & 6M & 9M & 1Y & 3Y \\
    \midrule
    Heston & 37.06 & 24.98 & 16.61 & 13.34 & 9.57 & 11.27 & 14.98 & 32.24 \\
    Heston w/ Jumps & 33.45 & 19.62 & 12.81 & 9.84 & 8.48 & 10.45 & 13.50 & 29.37 \\
    Heston++ & 21.40 & 9.73 & 9.81 & 8.58 & 9.38 & 10.71 & 12.58 & 26.65 \\
    $\sigma$-VAE w/o Residual & 22.63 & 19.61 & 14.81 & 12.84 & 9.76 & 9.72 & 13.18 & 27.36 \\
    $\sigma$-VAE w/ Residual & \textbf{7.73} & 6.18 & \textbf{5.25} & 5.14 & 3.90 & 3.78 & 5.98 & \textbf{10.87} \\
    $\Sigma$-VAE w/ Residual & 7.95 & \textbf{5.61} & 7.37 & \textbf{4.20} & \textbf{3.57} & \textbf{3.59} & \textbf{4.77} & 13.59 \\
    \bottomrule
    \end{tabular}
    \caption{USDCAD}
    \end{subtable}
    
    \begin{subtable}{\textwidth}
    \smaller
    \centering
    \begin{tabular}{lrrrrrrrr}
    \toprule
    Tenor & 1W & 1M & 2M & 3M & 6M & 9M & 1Y & 3Y \\
    \midrule
    Heston & 41.23 & 28.59 & 19.44 & 18.07 & 12.22 & 9.84 & 13.21 & 38.90 \\
    Heston w/ Jumps & 42.44 & 27.68 & 15.03 & 12.13 & 10.46 & 12.19 & 14.70 & 33.83 \\
    Heston++ & 21.67 & 12.39 & 9.51 & 9.33 & 10.95 & 11.33 & 10.96 & 22.06 \\
    $\sigma$-VAE w/o Residual & 24.83 & 25.40 & 15.77 & 11.64 & 10.79 & 11.24 & 11.58 & 25.85 \\
    $\sigma$-VAE w/ Residual & \textbf{10.65} & \textbf{8.73} & \textbf{6.48} & 5.58 & 5.20 & 3.69 & 5.64 & \textbf{9.05} \\
    $\Sigma$-VAE w/ Residual & 14.28 & 11.03 & 11.00 & \textbf{5.21} & \textbf{4.42} & \textbf{2.85} & \textbf{4.35} & 14.97 \\
    \bottomrule
    \end{tabular}
    \caption{GBPUSD}
    \end{subtable}
    
    \begin{subtable}{\textwidth}
    \smaller
    \centering
    \begin{tabular}{lrrrrrrrr}
    \toprule
    Tenor & 1W & 1M & 2M & 3M & 6M & 9M & 1Y & 3Y \\
    \midrule
    Heston & 63.24 & 28.90 & 32.48 & 29.94 & 19.30 & 16.61 & 18.75 & 80.49 \\
    Heston w/ Jumps & 45.38 & 30.54 & 22.47 & 18.43 & 15.54 & 20.09 & 23.41 & 54.06 \\
    Heston++ & 48.66 & 18.82 & 23.10 & 23.01 & 23.13 & 22.22 & 25.44 & 59.87 \\
    $\sigma$-VAE w/o Residual & 75.09 & 62.48 & 38.86 & 30.70 & 26.49 & 27.07 & 31.59 & 90.88 \\
    $\sigma$-VAE w/ Residual & \textbf{23.02} & 11.83 & \textbf{10.92} & \textbf{7.36} & \textbf{7.56} & 8.05 & \textbf{8.91} & \textbf{23.91} \\
    $\Sigma$-VAE w/ Residual & 29.85 & \textbf{9.37} & 11.88 & 9.07 & 9.99 & \textbf{7.81} & 11.75 & 41.53 \\
    \bottomrule
    \end{tabular}
    \caption{USDMXN}
    \end{subtable}
    
    \caption{Mean absolute error (MAE) measured in bps on the test set for varying tenors for different surfaces with missingness rate of 10\%.}
    \end{table}
    \begin{table}
    \begin{subtable}{\textwidth}
    \smaller
    \centering
    \begin{tabular}{lrrrrrrrr}
    \toprule
    Tenor & 1W & 1M & 2M & 3M & 6M & 9M & 1Y & 3Y \\
    \midrule
    Heston & 51.46 & 29.53 & 23.64 & 21.41 & 16.60 & 15.73 & 17.23 & 52.10 \\
    Heston w/ Jumps & 50.98 & 26.81 & 17.71 & 16.07 & 13.38 & 14.43 & 16.55 & 37.94 \\
    Heston++ & 31.76 & 19.55 & 15.33 & 13.93 & 14.21 & 15.43 & 17.87 & 44.63 \\
    $\sigma$-VAE w/o Residual & 30.65 & 23.75 & 19.57 & 18.72 & 16.22 & 14.94 & 18.68 & 33.39 \\
    $\sigma$-VAE w/ Residual & \textbf{18.59} & \textbf{14.78} & \textbf{11.62} & \textbf{9.42} & 6.43 & 6.23 & 8.87 & 10.94 \\
    $\Sigma$-VAE w/ Residual & 19.68 & 16.82 & 12.92 & 9.43 & \textbf{6.33} & \textbf{6.21} & \textbf{8.15} & \textbf{9.75} \\
    \bottomrule
    \end{tabular}
    \caption{AUDUSD}
    \end{subtable}
    
    \begin{subtable}{\textwidth}
    \smaller
    \centering
    \begin{tabular}{lrrrrrrrr}
    \toprule
    Tenor & 1W & 1M & 2M & 3M & 6M & 9M & 1Y & 3Y \\
    \midrule
    Heston & 39.13 & 27.59 & 19.31 & 16.46 & 12.34 & 14.09 & 18.81 & 30.94 \\
    Heston w/ Jumps & 51.39 & 25.91 & 16.31 & 13.83 & 12.06 & 15.56 & 18.74 & 33.12 \\
    Heston++ & 23.52 & 16.83 & 12.71 & 11.39 & 11.81 & 12.59 & 15.75 & 31.03 \\
    $\sigma$-VAE w/o Residual & 22.85 & 16.68 & 15.59 & 14.34 & 12.03 & 14.54 & 19.60 & 20.59 \\
    $\sigma$-VAE w/ Residual & \textbf{17.15} & \textbf{10.47} & \textbf{8.31} & 7.78 & 7.40 & 7.64 & \textbf{9.23} & \textbf{11.75} \\
    $\Sigma$-VAE w/ Residual & 20.62 & 12.49 & 10.30 & \textbf{7.45} & \textbf{6.80} & \textbf{7.11} & 9.74 & 13.51 \\
    \bottomrule
    \end{tabular}
    \caption{EURUSD}
    \end{subtable}
    
    \begin{subtable}{\textwidth}
    \smaller
    \centering
    \begin{tabular}{lrrrrrrrr}
    \toprule
    Tenor & 1W & 1M & 2M & 3M & 6M & 9M & 1Y & 3Y \\
    \midrule
    Heston & 39.76 & 25.45 & 17.53 & 14.36 & 10.70 & 12.36 & 17.16 & 36.11 \\
    Heston w/ Jumps & 51.70 & 25.52 & 14.57 & 11.81 & 11.04 & 13.18 & 16.95 & 33.54 \\
    Heston++ & 24.81 & 15.58 & 12.87 & 12.58 & 13.42 & 13.69 & 16.32 & 35.36 \\
    $\sigma$-VAE w/o Residual & 29.25 & 20.93 & 15.12 & 13.09 & 10.09 & 10.34 & 14.23 & 26.94 \\
    $\sigma$-VAE w/ Residual & 15.28 & 10.07 & \textbf{6.27} & 6.00 & 4.92 & 4.69 & 7.09 & \textbf{15.09} \\
    $\Sigma$-VAE w/ Residual & \textbf{14.72} & \textbf{9.76} & 7.82 & \textbf{5.34} & \textbf{4.37} & \textbf{4.19} & \textbf{5.87} & 15.20 \\
    \bottomrule
    \end{tabular}
    \caption{USDCAD}
    \end{subtable}
    
    \begin{subtable}{\textwidth}
    \smaller
    \centering
    \begin{tabular}{lrrrrrrrr}
    \toprule
    Tenor & 1W & 1M & 2M & 3M & 6M & 9M & 1Y & 3Y \\
    \midrule
    Heston & 43.10 & 29.41 & 19.60 & 17.94 & 12.91 & 12.94 & 17.69 & 38.62 \\
    Heston w/ Jumps & 51.50 & 28.59 & 16.72 & 14.81 & 12.37 & 14.56 & 18.47 & 36.94 \\
    Heston++ & 28.83 & 18.79 & 14.83 & 13.05 & 13.64 & 14.73 & 16.63 & 38.39 \\
    $\sigma$-VAE w/o Residual & 30.66 & 25.65 & 16.64 & 13.39 & 11.72 & 11.27 & 12.97 & 25.43 \\
    $\sigma$-VAE w/ Residual & \textbf{18.82} & \textbf{11.91} & \textbf{8.14} & 6.80 & 5.53 & 5.39 & 7.24 & \textbf{14.34} \\
    $\Sigma$-VAE w/ Residual & 23.47 & 15.02 & 10.49 & \textbf{6.67} & \textbf{4.97} & \textbf{4.40} & \textbf{5.22} & 17.26 \\
    \bottomrule
    \end{tabular}
    \caption{GBPUSD}
    \end{subtable}
    
    \begin{subtable}{\textwidth}
    \smaller
    \centering
    \begin{tabular}{lrrrrrrrr}
    \toprule
    Tenor & 1W & 1M & 2M & 3M & 6M & 9M & 1Y & 3Y \\
    \midrule
    Heston & 65.45 & 33.37 & 30.62 & 27.98 & 21.35 & 21.09 & 26.83 & 81.73 \\
    Heston w/ Jumps & 57.92 & 32.38 & 22.70 & 19.87 & 18.74 & 22.49 & 27.76 & 58.51 \\
    Heston++ & 55.77 & 28.46 & 26.02 & 26.94 & 26.66 & 25.99 & 32.15 & 80.67 \\
    $\sigma$-VAE w/o Residual & 80.32 & 59.49 & 40.63 & 30.97 & 26.70 & 28.82 & 35.07 & 92.66 \\
    $\sigma$-VAE w/ Residual & \textbf{28.10} & \textbf{13.41} & \textbf{12.46} & \textbf{10.02} & \textbf{8.66} & \textbf{9.00} & \textbf{11.06} & \textbf{36.74} \\
    $\Sigma$-VAE w/ Residual & 36.24 & 15.15 & 13.40 & 11.10 & 10.46 & 10.08 & 14.35 & 50.15 \\
    \bottomrule
    \end{tabular}
    \caption{USDMXN}
    \end{subtable}
    
    \caption{Mean absolute error (MAE) measured in bps on the test set for varying tenors for different surfaces with missingness rate of 50\%.}
    \end{table}
\begin{table}
\begin{subtable}{\textwidth}
\smaller
\centering
\begin{tabular}{lrrrrrrrr}
\toprule
Tenor & 1W & 1M & 2M & 3M & 6M & 9M & 1Y & 3Y \\
\midrule
Heston & 70.49 & 44.10 & 40.40 & 37.16 & 20.91 & 17.10 & 22.36 & 58.93 \\
Heston w/ Jumps & 57.36 & 39.65 & 32.85 & 28.39 & 18.25 & 20.33 & 25.86 & 41.10 \\
Heston++ & 40.85 & \textbf{21.56} & \textbf{16.07} & 19.79 & 19.98 & 19.05 & 19.37 & 47.46 \\
$\sigma$-VAE w/o Residual & 35.68 & 33.30 & 32.20 & 32.63 & 23.64 & 18.71 & 21.15 & 43.70 \\
$\sigma$-VAE w/ Residual & \textbf{25.22} & 21.67 & 19.68 & \textbf{13.23} & 11.35 & \textbf{9.03} & 13.10 & \textbf{13.76} \\
$\Sigma$-VAE w/ Residual & 50.49 & 24.74 & 27.54 & 16.48 & \textbf{9.56} & 10.92 & \textbf{11.73} & 15.78 \\
\bottomrule
\end{tabular}
\caption{AUDUSD}
\end{subtable}

\begin{subtable}{\textwidth}
\smaller
\centering
\begin{tabular}{lrrrrrrrr}
\toprule
Tenor & 1W & 1M & 2M & 3M & 6M & 9M & 1Y & 3Y \\
\midrule
Heston & 45.21 & 32.68 & 28.25 & 24.63 & 12.53 & 13.64 & 21.59 & 38.28 \\
Heston w/ Jumps & 38.44 & 29.54 & 22.68 & 19.24 & 12.05 & 16.13 & 19.79 & 28.14 \\
Heston++ & 18.74 & 13.83 & 13.13 & 11.99 & 10.01 & 12.65 & 14.66 & 30.65 \\
$\sigma$-VAE w/o Residual & 29.38 & 25.61 & 30.96 & 30.26 & 22.83 & 28.83 & 33.16 & 27.94 \\
$\sigma$-VAE w/ Residual & \textbf{15.21} & 11.04 & \textbf{11.44} & \textbf{11.39} & 8.61 & \textbf{6.74} & 9.55 & 15.27 \\
$\Sigma$-VAE w/ Residual & 17.08 & \textbf{9.67} & 19.59 & 11.81 & \textbf{7.18} & 7.00 & \textbf{7.95} & \textbf{15.08} \\
\bottomrule
\end{tabular}
\caption{EURUSD}
\end{subtable}

\begin{subtable}{\textwidth}
\smaller
\centering
\begin{tabular}{lrrrrrrrr}
\toprule
Tenor & 1W & 1M & 2M & 3M & 6M & 9M & 1Y & 3Y \\
\midrule
Heston & 45.23 & 30.57 & 26.18 & 21.03 & 12.32 & 15.43 & 23.95 & 43.47 \\
Heston w/ Jumps & 38.15 & 30.21 & 19.72 & 18.32 & 11.34 & 17.79 & 24.29 & 40.70 \\
Heston++ & 29.09 & 13.15 & 11.85 & 12.58 & 11.83 & 11.67 & 15.62 & 29.22 \\
$\sigma$-VAE w/o Residual & 31.52 & 30.21 & 27.58 & 24.10 & 18.35 & 20.06 & 26.15 & 27.41 \\
$\sigma$-VAE w/ Residual & \textbf{20.27} & 13.41 & \textbf{11.06} & 8.01 & 8.29 & 6.44 & 8.93 & \textbf{11.16} \\
$\Sigma$-VAE w/ Residual & 21.28 & \textbf{11.83} & 16.39 & \textbf{6.68} & \textbf{6.39} & \textbf{5.42} & \textbf{7.68} & 13.87 \\
\bottomrule
\end{tabular}
\caption{USDCAD}
\end{subtable}

\begin{subtable}{\textwidth}
\smaller
\centering
\begin{tabular}{lrrrrrrrr}
\toprule
Tenor & 1W & 1M & 2M & 3M & 6M & 9M & 1Y & 3Y \\
\midrule
Heston & 57.15 & 38.92 & 40.66 & 35.41 & 21.16 & 21.06 & 28.81 & 54.84 \\
Heston w/ Jumps & 44.67 & 35.14 & 33.06 & 26.02 & 20.12 & 22.01 & 26.43 & 42.75 \\
Heston++ & 33.78 & 19.13 & 18.39 & 15.59 & 14.32 & 16.19 & 14.95 & 27.86 \\
$\sigma$-VAE w/o Residual & 46.76 & 39.57 & 27.19 & 26.79 & 20.99 & 25.46 & 24.38 & 29.07 \\
$\sigma$-VAE w/ Residual & \textbf{19.99} & \textbf{15.33} & \textbf{14.15} & \textbf{12.02} & \textbf{8.93} & \textbf{7.37} & 12.55 & \textbf{17.41} \\
$\Sigma$-VAE w/ Residual & 41.70 & 17.19 & 23.46 & 12.89 & 12.10 & 8.96 & \textbf{9.76} & 23.38 \\
\bottomrule
\end{tabular}
\caption{GBPUSD}
\end{subtable}

\begin{subtable}{\textwidth}
\smaller
\centering
\begin{tabular}{lrrrrrrrr}
\toprule
Tenor & 1W & 1M & 2M & 3M & 6M & 9M & 1Y & 3Y \\
\midrule
Heston & 126.05 & 73.65 & 71.11 & 57.25 & 34.97 & 32.65 & 38.60 & 121.04 \\
Heston w/ Jumps & 76.56 & 71.93 & 56.72 & 41.39 & 30.52 & 35.93 & 45.07 & 117.06 \\
Heston++ & 79.24 & 49.86 & 47.41 & 46.57 & 44.34 & 39.33 & 42.85 & 77.41 \\
$\sigma$-VAE w/o Residual & 45.22 & 52.60 & 41.30 & 46.17 & 37.72 & 28.48 & 37.72 & 83.71 \\
$\sigma$-VAE w/ Residual & \textbf{17.55} & \textbf{20.48} & \textbf{16.27} & \textbf{15.52} & \textbf{13.34} & \textbf{13.18} & 15.15 & 56.29 \\
$\Sigma$-VAE w/ Residual & 181.64 & 59.46 & 51.45 & 27.44 & 20.56 & 13.41 & \textbf{14.27} & \textbf{44.97} \\
\bottomrule
\end{tabular}
\caption{USDMXN}
\end{subtable}

\caption{Mean absolute error (MAE) measured in bps on the validation set for varying tenors for different surfaces with missingness rate of 10\%.}
\end{table}
\begin{table}
\begin{subtable}{\textwidth}
\smaller
\centering
\begin{tabular}{lrrrrrrrr}
\toprule
Tenor & 1W & 1M & 2M & 3M & 6M & 9M & 1Y & 3Y \\
\midrule
Heston & 75.35 & 47.54 & 44.11 & 39.46 & 25.97 & 23.19 & 30.43 & 63.74 \\
Heston w/ Jumps & 78.40 & 50.63 & 38.56 & 31.33 & 22.13 & 23.31 & 29.28 & 50.25 \\
Heston++ & 49.02 & 30.05 & 23.52 & 23.95 & 23.39 & 22.69 & 26.14 & 64.46 \\
$\sigma$-VAE w/o Residual & 42.21 & 36.02 & 35.43 & 39.95 & 25.02 & 20.29 & 27.46 & 46.11 \\
$\sigma$-VAE w/ Residual & \textbf{33.79} & \textbf{27.66} & \textbf{20.71} & \textbf{16.58} & 12.95 & \textbf{9.91} & 13.92 & \textbf{16.37} \\
$\Sigma$-VAE w/ Residual & 55.41 & 33.05 & 30.98 & 19.09 & \textbf{12.31} & 10.13 & \textbf{13.79} & 18.51 \\
\bottomrule
\end{tabular}
\caption{AUDUSD}
\end{subtable}

\begin{subtable}{\textwidth}
\smaller
\centering
\begin{tabular}{lrrrrrrrr}
\toprule
Tenor & 1W & 1M & 2M & 3M & 6M & 9M & 1Y & 3Y \\
\midrule
Heston & 49.84 & 33.03 & 29.23 & 24.10 & 14.50 & 15.72 & 21.98 & 40.70 \\
Heston w/ Jumps & 58.91 & 32.78 & 23.90 & 19.03 & 14.50 & 15.98 & 20.69 & 35.39 \\
Heston++ & 28.23 & 18.71 & 14.35 & 13.95 & 13.90 & 14.88 & 18.48 & 41.09 \\
$\sigma$-VAE w/o Residual & 32.09 & 28.05 & 32.25 & 30.68 & 25.77 & 29.83 & 36.15 & 29.01 \\
$\sigma$-VAE w/ Residual & \textbf{18.18} & \textbf{14.98} & \textbf{13.69} & \textbf{12.83} & 11.15 & \textbf{10.12} & 12.53 & \textbf{17.40} \\
$\Sigma$-VAE w/ Residual & 20.38 & 15.61 & 22.10 & 13.42 & \textbf{10.78} & 10.32 & \textbf{11.27} & 18.26 \\
\bottomrule
\end{tabular}
\caption{EURUSD}
\end{subtable}

\begin{subtable}{\textwidth}
\smaller
\centering
\begin{tabular}{lrrrrrrrr}
\toprule
Tenor & 1W & 1M & 2M & 3M & 6M & 9M & 1Y & 3Y \\
\midrule
Heston & 45.84 & 32.92 & 28.89 & 23.34 & 14.88 & 17.44 & 24.38 & 43.89 \\
Heston w/ Jumps & 60.08 & 33.98 & 24.77 & 19.65 & 16.00 & 20.32 & 26.54 & 41.99 \\
Heston++ & 34.77 & 21.56 & 15.34 & 16.15 & 15.65 & 15.38 & 19.62 & 44.34 \\
$\sigma$-VAE w/o Residual & 34.60 & 30.57 & 27.60 & 26.75 & 20.60 & 22.60 & 31.65 & 31.53 \\
$\sigma$-VAE w/ Residual & 26.61 & 18.53 & \textbf{14.47} & 11.67 & 9.87 & 8.20 & 10.83 & 16.87 \\
$\Sigma$-VAE w/ Residual & \textbf{25.88} & \textbf{18.08} & 19.76 & \textbf{10.36} & \textbf{8.11} & \textbf{6.78} & \textbf{9.13} & \textbf{16.60} \\
\bottomrule
\end{tabular}
\caption{USDCAD}
\end{subtable}

\begin{subtable}{\textwidth}
\smaller
\centering
\begin{tabular}{lrrrrrrrr}
\toprule
Tenor & 1W & 1M & 2M & 3M & 6M & 9M & 1Y & 3Y \\
\midrule
Heston & 63.85 & 42.03 & 39.92 & 35.13 & 23.63 & 23.53 & 31.88 & 60.10 \\
Heston w/ Jumps & 63.72 & 42.09 & 32.74 & 27.21 & 23.75 & 26.03 & 31.14 & 52.52 \\
Heston++ & 45.93 & 25.85 & 20.63 & 19.52 & 19.37 & 20.11 & 23.80 & 50.17 \\
$\sigma$-VAE w/o Residual & 56.25 & 40.79 & 28.58 & 27.62 & 24.39 & 24.43 & 26.05 & 30.01 \\
$\sigma$-VAE w/ Residual & \textbf{25.25} & \textbf{19.88} & \textbf{15.00} & \textbf{13.76} & \textbf{10.10} & \textbf{9.24} & \textbf{11.68} & \textbf{22.55} \\
$\Sigma$-VAE w/ Residual & 48.52 & 22.63 & 25.20 & 15.86 & 13.01 & 11.20 & 12.44 & 27.06 \\
\bottomrule
\end{tabular}
\caption{GBPUSD}
\end{subtable}

\begin{subtable}{\textwidth}
\smaller
\centering
\begin{tabular}{lrrrrrrrr}
\toprule
Tenor & 1W & 1M & 2M & 3M & 6M & 9M & 1Y & 3Y \\
\midrule
Heston & 130.17 & 88.69 & 81.05 & 66.53 & 42.46 & 38.66 & 45.53 & 114.54 \\
Heston w/ Jumps & 106.10 & 79.75 & 60.69 & 48.99 & 35.86 & 41.74 & 52.26 & 107.66 \\
Heston++ & 97.97 & 74.02 & 55.58 & 51.75 & 49.03 & 47.32 & 54.64 & 111.72 \\
$\sigma$-VAE w/o Residual & 57.03 & 52.72 & 47.21 & 50.14 & 41.49 & 34.62 & 43.04 & 85.54 \\
$\sigma$-VAE w/ Residual & \textbf{41.82} & \textbf{32.94} & \textbf{22.79} & \textbf{17.83} & \textbf{15.82} & \textbf{12.97} & \textbf{16.03} & 53.01 \\
$\Sigma$-VAE w/ Residual & 195.80 & 89.46 & 69.56 & 38.89 & 26.97 & 17.96 & 20.08 & \textbf{49.63} \\
\bottomrule
\end{tabular}
\caption{USDMXN}
\end{subtable}

\caption{Mean absolute error (MAE) measured in bps on the validation set for varying tenors for different surfaces with missingness rate of 50\%.}
\end{table}

\end{document}